\documentclass[traditabstract]{aa} % for the abstract without structuration % (traditional abstract) 
\usepackage{graphicx}
\usepackage{txfonts}
\usepackage{natbib}
\usepackage{longtable}
\def\ms{\hbox{\,m\,s$^{-1}$}}         %m.s -1
\def\m2s2{\hbox{\,m$^{2}$\,s$^{-2}$}} %m2.s -2
\def\kms{\hbox{\,km\,s$^{-1}$}}       %km.s -1
\def\Msun{\hbox{$\mathrm{M}_{\odot}$}}             %Msun
\def\Rsun{\hbox{$\mathrm{R}_{\odot}$}}
\def\Mjup{\hbox{$\mathrm{M}_{\rm Jup}$}}
\def\Rjup{\hbox{$\mathrm{R}_{\rm Jup}$}}

\def\mp{$M_{\rm p}$}
\def\rp{$R_{\rm p}$}

\begin{document}

\title{SOPHIE velocimetry of \emph{Kepler} transit candidates\thanks{Based on observations made with SOPHIE on the 1.93-m telescope
at Observatoire de Haute-Provence (CNRS), France}}
\subtitle{V. The three hot Jupiters KOI-135b, KOI-204b and KOI-203b (alias Kepler-17b)}

\titlerunning{The three hot Jupiters KOI-135b, KOI-204b and KOI-203b (alias Kepler-17b)}
\authorrunning{Bonomo et al. 2011}

\author{A.~S.~Bonomo \inst{1} 
\and G.~H\'ebrard \inst{2, 3} 
\and A.~Santerne\inst{1, 3}
\and N.~C.~Santos \inst{4, 5}
\and M.~Deleuil \inst{1}
\and J.~Almenara \inst{1}
\and F.~Bouchy \inst{2, 3}  
\and R.~F.~D\'iaz\inst{2, 3}
\and C.~Moutou\inst{1} 
\and M.~Vanhuysse \inst{6}.
}
%\and G.~H\'ebrard \inst{2, 3}
%\and C.~Moutou\inst{1} 
%\and R.~D\'iaz\inst{2, 3}
%\and A.~Bouchy \inst{2, 3}  
%\and A.~Santerne\inst{1, 3}
%\and M.~Deleuil \inst{1}
%\and J.~Almenara \inst{1}
% \and Oversky, 47 AllŽe des Palanques, 33127, Saint Jean d'Illac, France}
%\and \textbf{anybody else I forgot?}}	

\institute{
Laboratoire d'Astrophysique de Marseille, Universit\'e Aix-Marseille \& CNRS, 38 rue Fr\'ed\'eric Joliot-Curie, F-13388 Marseille Cedex 13, France
\and Institut d'Astrophysique de Paris, UMR7095 CNRS, Universit\'e Pierre \& Marie Curie, 98bis boulevard Arago, 75014 Paris, France
\and Observatoire de Haute-Provence, Universit\'e Aix-Marseille \& CNRS, F-04870 St.~Michel l'Observatoire, France
\and Centro de Astrof{\'\i}sica, Universidade do Porto, Rua das Estrelas, 4150-762 Porto, Portugal
\and Departamento de F{\'\i}sica e Astronomia, Faculdade de Ci\^encias, Universidade do Porto, Portugal
\and Oversky, 47 AllŽe des Palanques, 33127, Saint Jean d'Illac, France
}

\date{Received ?? / Accepted ??}

\offprints{\\
 \email{aldo.bonomo@oamp.fr}}

\abstract{We report the discovery of two new transiting hot Jupiters, KOI-135b and KOI-204b, that were previously identified
as planetary candidates by Borucki et al. 2011, and, independently of the Kepler team, confirm the planetary nature of Kepler-17b, 
recently announced by D\'esert et al. 2011. Radial-velocity measurements, taken
with the SOPHIE spectrograph at the Observatoire de Haute-Provence (France), and Kepler photometry (Q1 and Q2 data) were
used to derive the orbital, stellar and planetary parameters. 
KOI-135b and KOI-204b orbit their parent stars in $\sim 3.02$ and $3.25$~days, 
respectively.  They have approximately the same radius, \rp~$=1.20 \pm 0.06$~\Rjup ~and $1.24 \pm 0.07$~\Rjup, 
but different masses \mp~$=3.23 \pm 0.19$~\Mjup ~and $1.02 \pm 0.07$~\Mjup. As a consequence, their bulk 
densities differ by a factor of four, $\rho_{\rm p}=2.33 \pm 0.36\, \rm g\,cm^{-3} $ (KOI-135b)~and $0.65 
\pm 0.12\, \rm g\,cm^{-3}$ (KOI-204b), meaning that 
their interior structures are different. All the three planets orbit metal-rich stars with 
$[\rm{Fe/H}] \sim 0.3$~dex. Our SOPHIE spectra of Kepler-17b, used both to measure 
the radial-velocity variations and determine the atmospheric parameters of the host star, 
allow us to refine the characterisation of the planetary system. In particular we found the
radial-velocity semi-amplitude and the stellar mass to be respectively slightly smaller 
and larger than D\'esert et al. These two quantities, however, 
compensate and lead to a planetary mass fully consistent with D\'esert et al. : 
our analysis gives \mp~$=2.47 \pm 0.10$~\Mjup ~and \rp~$=1.33 \pm 0.04$~\Rjup.
We found evidence for a younger age of this planetary system, $t <1.8$~Gyr, which is supported 
by both evolutionary tracks and gyrochronology. 
Finally, we confirm the detection of the optical secondary eclipse by D\'esert et al.   
and found also the brightness phase variation with the Q1 and Q2 Kepler data. The latter indicates a low redistribution 
of stellar heat to the night side ($<16 \%$ at 1-$\sigma$), if the optical 
planetary occultation comes entirely from thermal flux. The geometric albedo is $A_{\rm g}<0.12$ (1-$\sigma$).}
%{To be done}
%{To be done}
%{To be done}
%{To be done}

\keywords{planetary systems -- stars: fundamental parameters -- 
techniques: photometric -- techniques: spectroscopic -- techniques: radial velocities.}

% 5 {} token are mandatory

\maketitle

%
%________________________________________________________________

%	\abstract{CoRoT is the pioneer space mission dedicated to the detection of extrasolar
%planets via the transit method. It continueously monitors the optical flux of thousands of stars
%in each of its fields of view.}


\section{Introduction}
\noindent
Since two years and a half, the Kepler space telescope is monitoring
the optical flux of about 156,000 stars with $\rm 9 < V < 16$ in the Cygnus
constellation to search for transiting planets (e.g., \citealt{Boruckietal06}).
More than 1200 planetary candidates have been announced 
by \citet{Boruckietal11} in February 2011. However, all of them cannot be followed 
up by the Kepler team as this would require a huge amount of telescope time.
Since the Kepler team publicly announced all the 
planetary candidates they found, this allows also other groups to carry out 
spectroscopic observations of their targets. 

%SOPHIE \citep{bouchy09} is a cross-dispersed, environmentally 
%stabilized, fiber-fed, echelle spectrographs dedicated to high-precision radial velocity 
%measurements.

The SOPHIE spectrograph \citep{bouchy09}, 
mounted on the 1.93~m telescope at the Observatoire de Haute Provence 
(France), has been performing very well to assess the nature of CoRoT 
planetary candidates and derive the orbital parameters of many CoRoT planets
(e.g., \citealt{Deleuiletal11}). Therefore, about one year and a half ago, we 
started the spectroscopic follow-up of Kepler candidates 
with Jupiter and Saturn sizes, around stars with 
$\rm K_{p}$\footnote{Kepler magnitude}$< 14.7$. So far, this follow-up campaign led 
to several interesting discoveries: one new hot Jupiter orbiting an 
evolved star, KOI-428b \citep{Santerneetal10}; KOI-423b, a 18~\Mjup ~companion 
orbiting a subgiant F7IV star, that could be either a
low-mass brown dwarf or an extremely massive planet \citep{Bouchyetal11}; and
the hot Jupiter KOI-196b for which we detected the optical secondary eclipse and phase
variations \citep{Santerneetal11}. 
Moreover, this follow-up campaign is revealing the rate of Kepler 
Jupiter-size planetary candidates that are actually false positives,
to be compared with the $\sim 8\%$ upper limit predicted by 
\citet{MortonJohnson11} (Santerne et al., in preparation).

%The estimated geometric albedo of KOI-196b,
%$A_{g}=0.30 \pm 0.08$, is significantly higher than predicted by models of 
%planetary atmospheres and, thus, together with Kepler-7b (Demory et al. 2011, ADD ref.)
%and HAT-P-7b (Welsh et al., ADD ref.), it represents a test bench for such models.

In the present paper, we announce the discovery of two new hot Jupiters with an 
orbital period of $\sim 3$~days, namely KOI-135b and KOI-204b.
Orbital and planetary parameters for these two planets were derived
by performing a simultaneous
modelling of Kepler photometry and SOPHIE radial-velocity observations. 
Moreover, independently from the Kepler team, we confirm 
the planetary nature of Kepler-17b, a hot Jupiter around an active solar-like star, recently announced
by \citet{desert11}. Our new SOPHIE spectra, used both to measure the radial-velocity
variations and characterise the host star,
allow us to improve the parameters of this planetary system.
Besides that, we detected the optical phase variations of Kepler-17b, i.e.
the variation in brightness as the dayside of the planet rotates
into and out of view, not reported previously, 
and discuss its possible implications for the planet atmosphere.
%eclipse of Kepler-17b 
%in the Kepler light curve 

\section{KOI-135}
\subsection{Kepler observations}
\label{Kepler_obs_KOI135}
\begin{figure}[h]
\centering
\includegraphics[width=6.5cm, angle=90]{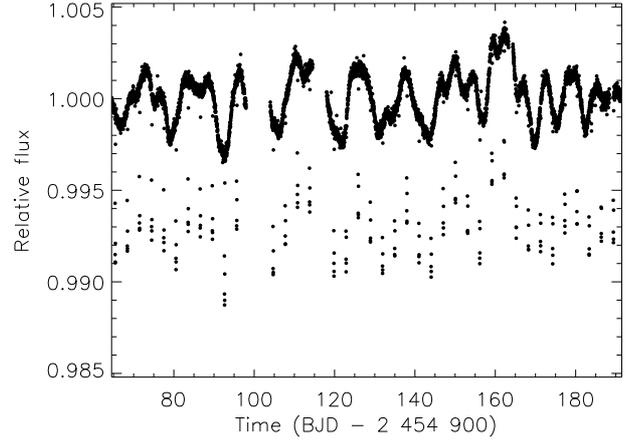}
\caption{The Kepler light curve of KOI-135 with a temporal sampling of 29.4~min
showing 39 transits and flux variations due to stellar activity with a peak-to-peak 
amplitude of $\sim 0.5$\%. Long-term trends and the steep variations after the safe 
modes were removed.}
\label{lcfig135}
\end{figure}

\noindent
KOI-135 is a $\rm K_{p}=14.0$ 
star and was observed by 
Kepler with a temporal sampling of 29.4 min (Long Cadence 
data) for 122.2 days: from May, 13 to June, 15, 2009 during 
the first 33.5-day segment of science operations (Q1), and from 
June, 20 to September, 16 of the same year, during the second 
quarter (Q2) data lasting 88.7 days. Its coordinates, magnitudes 
and IDs are listed in Table~\ref{startable_KOI}. The raw Kepler 
light curve, publicly available at the MAST 
archive\footnote{http://archive.stsci.edu/kepler/data\_search/search.php}, 
contains 5722 validated photometric measurements.
From these data points we subtracted the flux excess due to 
background stars contaminating the photometric mask, as estimated
by the Kepler team: 2.4\% for the Q1 and 3.1\% for the Q2 data
\footnote{http://archive.stsci.edu/kepler/kepler\_fov/search.php}.
%Long-term trends of clear instrumental origin and 
%the steep variations after the two safe modes 
%\citep{Jenkinsetal10} were then removed.
Long-term trends of clear instrumental origin and 
the steep variations after the two safe modes 
\citep{Jenkinsetal10} were then removed.
The so-treated light curve is shown in Fig.~\ref{lcfig135}. It exhibits thirty-nine
transits with a period of 3.02 days, a depth of 0.8\% and a duration 
of $\sim3$~hours. The standard deviation of the light curve, computed in a 
robust way after removing low frequency variations with a 
sliding median filter, is $180$~ppm, which is compatible
with the median of the errors of the single photometric measurements, 
i.e. $160$~ppm.

Stellar variability due to the rotational modulation of active regions
on the stellar photosphere is clearly visible in the light curve (Fig.~\ref{lcfig135}). 
Its peak-to-peak amplitude is about 0.5\%. After removing transits, 
we performed a periodogram analysis of
the light curve and computed the autocorrelation function in order to estimate
the stellar rotation period. The two methods give very consistent results: 
$P_{\rm *,rot}=12.9 \pm 0.7$~days.

%%%%%%%%%%%%%%%%%%%
%%%%%%%%%%%%%%%%%%%%
%%%%%%%%%%%%%%%%%%%
\begin{table*}
\centering
\caption{IDs, coordinates, and magnitudes of the planet-hosting stars KOI-135, 204, and 203.}            
%\begin{minipage}{3.5 cm} 
%\centering        
%\begin{minipage}[!]{10.0cm}  
\renewcommand{\footnoterule}{}     
\begin{tabular}{lcccc}       
\hline\hline                 
%BJD & RV & $\pm$$1\,\sigma$ & exp. time & S/N p. pix. \\
%-2\,400\,000 & (km\,s$^{-1}$) & (km\,s$^{-1}$) & (sec) &  (at 550 nm)  \\

Kepler Object of Interest & KOI-135 & KOI-204 & KOI-203 \\
~\\
Kepler ID & 9818381 & 9305831 & 10619192\\
USNO-A2 ID  & 1350-10117895 & 1350-11449251 & 1350-11245067\\
2MASS ID   & 19005780+4640057  & 20002456+4545437 & 19533486+4748540\\
%GSC2.3 ID & NIMR021985 \\
\\
\multicolumn{2}{l}{Coordinates} \\
\hline            
RA (J2000)   & 19:00:57.82  & 20:0:24.55 & 19:53:34.87 \\
Dec (J2000) &  46:40:5.88   & 45:45:43.56  & 47:48:54.0 \\
\\
\multicolumn{3}{l}{Magnitudes} \\
\hline
\centering
Filter &  &  \\
\hline
%B$^a$  & 16.68 & 0.14 \\
%V$^a$  & 15.22 & 0.05 \\
%B$^a$ & 14.6 & \\ 
%R$^a$ & 14.0 & \\
$ \rm K_{p}$$^a$ & 13.96 &  14.68 & 14.14\\
%r$^a$ & 13.88 &  \\
J$^b$  & 12.86 (0.02)  & 13.34 (0.03) & 12.99 (0.02) \\ 
H$^b$  & 12.60 (0.02) & 12.97 (0.02) & 12.67 (0.02) \\
K$^b$  & 12.55 (0.03) & 12.89 (0.03) & 12.58 (0.02) \\
%\\                                    
%\multicolumn{3}{l}{Proper motion} \\
%\hline
%$\mu_{\alpha}$ & 8.0   & ''/yr
%$\mu_{\alpha}$ & -11.8 & ''/yr
\hline\hline
%\vspace{-0.5cm}
%\footnotetext[1]{from USNO-A2.0 catalogue;}
%\footnotetext[1]{\scriptsize Kepler magnitude from MAST Archive;}
%\footnotetext[2]{\scriptsize from 2MASS catalogue.}
\multicolumn{4}{l}{$a$: Kepler magnitude from MAST Archive; $b$: from 2MASS catalogue.} \\
%\multicolumn{4}{l}{a: Kepler magnitude from MAST Archive;} 
%\multicolumn{4}{l}{b: from 2MASS catalogue.} \\
\end{tabular}
%\end{minipage}
\label{startable_KOI}      
\end{table*}
~\\
%%%%%%%%%%%%%%%%%%%
%%%%%%%%%%%%%%%%%%%%
%%%%%%%%%%%%%%%%%%%

%\begin{table}
%\caption{KOI-135 IDs, coordinates, and magnitudes.}            
%%\begin{minipage}{3.5 cm} 
%\centering        
%\begin{minipage}[!]{7.0cm}  
%\renewcommand{\footnoterule}{}     
%\begin{tabular}{lcc}       
%\hline\hline                 
%%BJD & RV & $\pm$$1\,\sigma$ & exp. time & S/N p. pix. \\
%%-2\,400\,000 & (km\,s$^{-1}$) & (km\,s$^{-1}$) & (sec) &  (at 550 nm)  \\
%Kepler ID & 9818381 \\
%USNO-A2 ID  & 1350-10117895 \\
%2MASS ID   & 19005780+4640057 \\
%GSC2.3 ID & NIMR021985 \\
%\\
%\multicolumn{2}{l}{Coordinates} \\
%\hline            
%RA (J2000)  & 19:00:57.82 \\
%Dec (J2000) & 46:40:5.88 \\
%\\
%\multicolumn{3}{l}{Magnitudes} \\
%\hline
%\centering
%Filter & Mag & Error \\
%\hline
%%B$^a$  & 16.68 & 0.14 \\
%%V$^a$  & 15.22 & 0.05 \\
%%B$^a$ & 14.6 & \\ 
%%R$^a$ & 14.0 & \\
%$ \rm K_{p}$$^a$ & 13.96 &  \\
%%r$^a$ & 13.88 &  \\
%J$^b$  & 12.86 & 0.02 \\ 
%H$^b$  & 12.60 & 0.02 \\
%K$^b$  & 12.55 & 0.03 \\
%%\\                                    
%%\multicolumn{3}{l}{Proper motion} \\
%%\hline
%%$\mu_{\alpha}$ & 8.0   & ''/yr
%%$\mu_{\alpha}$ & -11.8 & ''/yr
%\hline\hline
%\vspace{-0.5cm}
%%\footnotetext[1]{from USNO-A2.0 catalogue;}
%\footnotetext[1]{Kepler magnitude from MAST Archive;}
%\footnotetext[2]{from 2MASS catalogue.}
%\end{tabular}
%\end{minipage}
%\label{startable_KOI135}      
%\end{table}

%\begin{figure}[h]
%\centering
%\includegraphics[width=6.5cm, angle=90]{plot_lc_KOI423_zoom.ps}
%\caption{A zoom of the KOI-423 light curve showing one transit and stellar 
%variability.}
%\label{lcfig_zoom}
%\end{figure}

\subsection{Centroid analysis}
%To reject all background eclipsing binary located within the \textit{Kepler} 
%photometric mask which can mimic the observed transit, we analyzed the centroid motion measured by \textit{Kepler} during the transit \citep{batahlaetal2010}. Fig. \ref{rain_plot} shows the X \& Y centroid shifts for KOI-204 and KOI-135 after correction of any local variations of the LC and the centroid curve. KOI-135 didn't show any significant shift during the transit up to 0.07 mpixel. According to the MAST database, there is only one contaminant, KIC9818377, a Kp=16.5 located at 15.6\arcsec. Since there is no significant shift in the centroid, we assumed that KOI-135 is the transit host star. 

\begin{figure}[h]
\begin{center}
\includegraphics[width=\columnwidth]{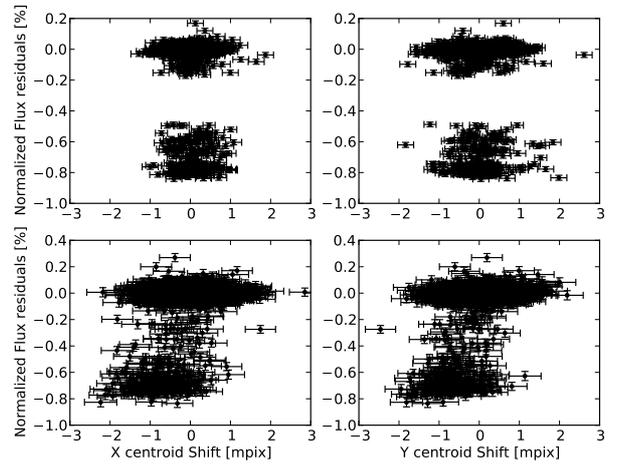}
\caption{Rain plots showing X (left panels) and Y (right panels) shifts of the centroid 
as a function of the normalized flux measured during the transits 
of KOI-135b (upper panels) and KOI-204b (lower panels).}
\label{rain_plot}
\vspace{-0.5cm}
\end{center}
\end{figure}

\noindent
In order to reject the scenario of a background eclipsing binary located within the
Kepler photometric mask, that might mimic the observed transits, the centroid
timeseries was analysed \citep{Batalhaetal2010}. KOI-135 does not show any significant centroid shift 
during the transit up to 0.07 mpixel (Fig.~\ref{rain_plot}). 
%According to the MAST database, there is only one contaminant with $Kp=16.5$, 
%KIC9818377, located at 15.6\arcsec. 

\subsection{Ground-based follow-up}

\subsubsection{Radial-velocity observations}
\label{sect_RV_KOI135}
%\textbf{TBD by Guillaume.}
\noindent
Eight spectra of KOI-135 were secured during the summer 2011 with the 
SOPHIE spectrograph.
%at the 1.93-m telescope of Haute-Provence Observatory, France. 
%SOPHIE \citep{bouchy09} is a cross-dispersed, environmentally 
%stabilized, fiber-fed, echelle spectrographs dedicated to high-precision radial velocity 
%measurements. 
They were acquired in High-Efficiency mode (resolution 
power $R=39\,000$ at 5500 \AA) and slow-mode for the reading of the detector. The first 
aperture fiber was put on the target whereas the second 2' away on 
the sky in order to evaluate the background.
The spectra extraction was performed using the SOPHIE pipeline. 
Following the techniques described by \citet{baranne96} and  
\citet{pepe02}, the radial velocities were measured from a 
weighted cross-correlation of the spectra with a numerical mask. 
We used a standard G2 mask that includes more than 3500 lines.
The resulting cross-correlation functions (CCFs) were fitted by Gaussians to get the 
radial velocities and the associated photon-noise errors. The full width at half 
maximum of those Gaussians is $11.6 \pm 0.1$~km\,s$^{-1}$, and its contrast is 
$24 \pm 2$~\%\ of the continuum.
We only used the spectral orders 12 to 38 in the cross-correlation 
to reduce the dispersion of the measurements produced by
noisy spectral orders. 
%Indeed, some spectral domains  
%are noisy, so using them degrades the accuracy of the radial velocity measurement. 
Moonlight contamination was tenuous in the spectra and required a correction only for two over 
the eight spectra, following the method described in 
\citet{pollacco08} and \citet{hebrard08}. It introduced 
modest corrections (7 and 20$\,\ms$ for the two spectra).

The log of the observations and the radial-velocity measurements are reported in 
Table~\ref{table_rv}. Radial-velocity precision ranges between 
12 and 29\,\ms\ depending on the exposure time and weather conditions. 
Table~\ref{table_rv} also lists the bisector spans that we measured on 
the cross-correlation functions in order to quantify the possible shape variations 
of the spectral lines. The error bars on the bisector spans were assumed to be 
twice those of the corresponding radial velocity.

\begin{figure*}[t] 
%\vspace{2. cm}
\begin{center}
\includegraphics[scale=0.5]{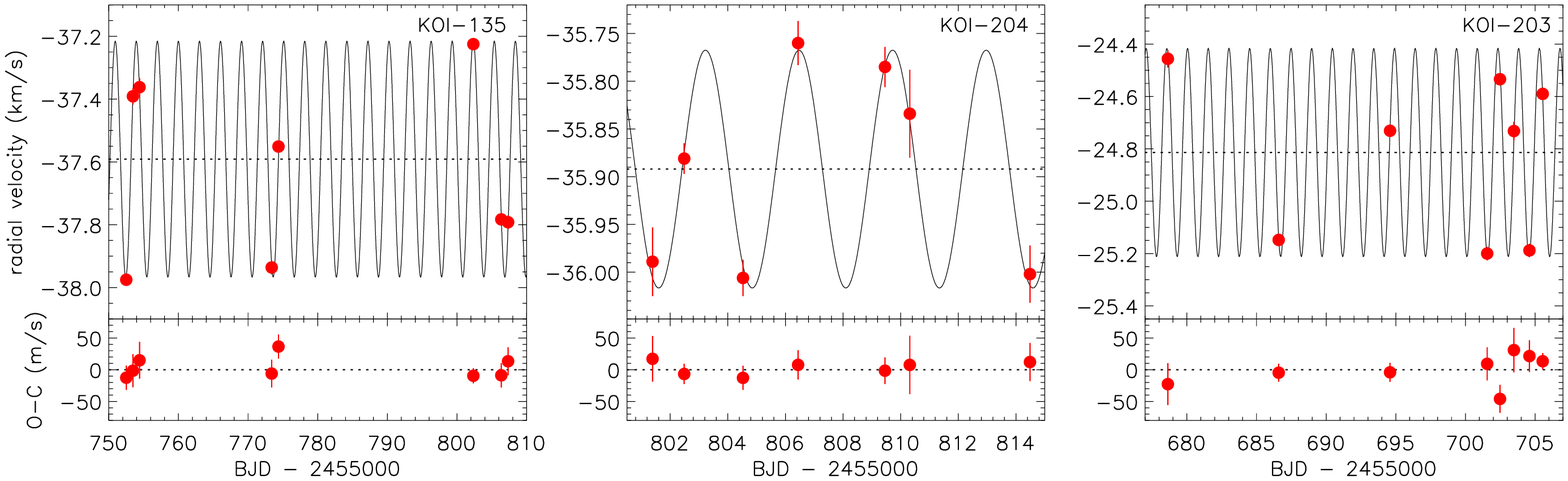}
\includegraphics[scale=0.5]{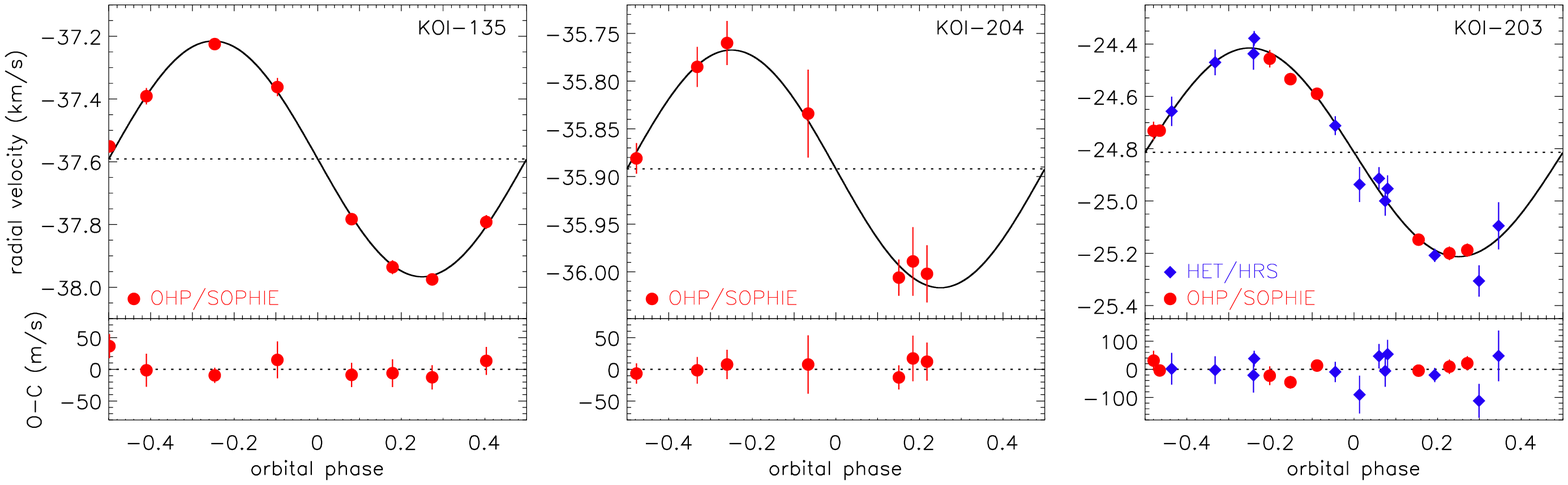}
\caption{\textit{Upper panel:} Radial-velocity measurements of the three targets
with 1-$\sigma$\,Êerror bars as a function of time together with their Keplerian 
fit (\textit{top}) and residuals of the fit. 
\textit{Lower panel:} 
Same as above but as a function of the orbital phase. 
}
\label{fig_orb_rv}
\end{center}
\end{figure*}

The measurements are displayed in Fig.~\ref{fig_orb_rv}, left panel, together with their 
circular Keplerian fit obtained from the global photometry and radial-velocity best fit 
(Sect.~\ref{system_par_KOI135}).
%assuming the period and transit epoch determined by the Kepler 
%light curve. 
The radial velocities present a semi-amplitude $K=375\pm13$~m\,s$^{-1}$, 
in phase with the Kepler ephemeris.
%corresponding to a planet with a mass 
%\mp~$  = 3.25 \pm 0.19$~\Mjup. This assumes $M_\star = 1.34\pm0.09$\,M$_\odot$ 
%for the host star, which here is the main source of uncertainty on~\mp.
The standard deviation of the residuals to the fit is $\sigma_{O-C}=15.9$~m\,s$^{-1}$. 
The reduced $\chi^2$ is 1.0 for the eight radial velocities used 
in the fit. We do not detect any drift over the 55-day span of the radial velocity.

\begin{figure*}[t] 
%\vspace{2. cm}
\begin{center}
\includegraphics[scale=0.5]{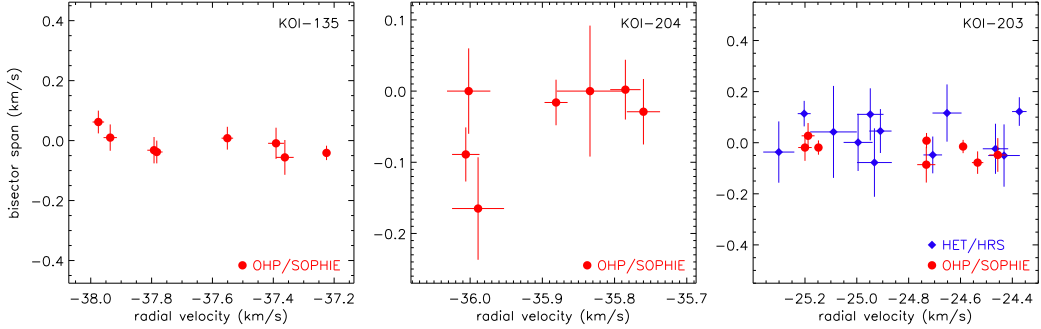}
\caption{Bisector span as a function of the radial velocities with 1-$\sigma$\,Êerror bars
for the three targets. 
The ranges have the same extents in the $x$- and $y$-axes.}
\label{fig_bis}
\end{center}
\end{figure*}

SOPHIE radial velocities obtained with different stellar masks (F0, K0, or K5) produce 
variations with the same amplitude as for the G2 mask and, thus, there is no 
indication for blend scenarios implying stars of different spectral types. Similarly, the CCF bisector
spans show neither variations nor trend as a function of radial velocity 
(Fig.~\ref{fig_bis}, left panel).
This reinforces the conclusion that the radial-velocity variations are caused by a 
planetary companion and not by changes in the 
spectral line profile due to blends.

\begin{table*}
\centering 
\caption{SOPHIE radial-velocity measurements.}
\begin{tabular}{lcccccc}
\hline
\hline
BJD$_{\rm UTC}$ & RV & $\pm$$1\,\sigma$ & Bis. span & exp. time & S/N p. pix. & Target \\
-2\,400\,000 & (km\,s$^{-1}$) & (km\,s$^{-1}$) & (km\,s$^{-1}$) & (sec) &  (at 550 nm) & \\
\hline
55752.5089                      &  -37.975  &  0.019  &   0.062  &  1562  &  19.5  &  KOI-135 \\  
55753.4652$^\dagger$ &  -37.391  &  0.026  &  -0.009  &  3600  &  16.0  &  KOI-135 \\     
55754.4127$^\dagger$ &  -37.362  &  0.029  &  -0.056  &  3600  &  14.2  &  KOI-135 \\
55773.3908                      &  -37.936  &  0.022  &   0.010  &  3600  &  17.5  &  KOI-135 \\
55774.3658                      &  -37.551  &  0.019  &   0.008  &  3600  &  18.4  &  KOI-135 \\
55802.3448                      &  -37.225  &  0.012  &  -0.041  &  3600  &  26.3  &  KOI-135 \\
55806.3595                      &  -37.783  &  0.019  &  -0.038  &  2703  &  19.0  &  KOI-135 \\
55807.3341                      &  -37.792  &  0.022  &  -0.032  &  3600  &  17.7  &  KOI-135 \\
\hline
55801.3890                      &  -35.989  &  0.036  &  -0.165  &  3600  &  10.9  &  KOI-204 \\         
55802.4872                      &  -35.881  &  0.016  &  -0.016  &  3600  &  18.3  &  KOI-204 \\          
55804.5266                      &  -36.006  &  0.019  &  -0.089  &  3101  &  16.0  &  KOI-204 \\    
55806.4383                      &  -35.760  &  0.023  &  -0.029  &  2249  &  14.1  &  KOI-204 \\
55809.4533                      &  -35.785  &  0.021  &   0.002  &  3600  &  14.9  &  KOI-204 \\ 
55810.3153$^\dagger$ &  -35.834  &  0.046  &  -0.000  &  2543  &  10.0  &  KOI-204 \\ 
55814.4852                      &  -36.002  &  0.030  &   0.000  &  3600  &  17.5  &  KOI-204 \\
\hline
55678.6344$^\dagger$ &  -24.456  &  0.033  &  -0.048  &  1562  &  12.3  &  KOI-203 \\          
55686.5921                      &  -25.148  &  0.014  &  -0.019  &  3600  &  22.8  &  KOI-203 \\     
55694.5860                      &  -24.731  &  0.015  &   0.008  &  3600  &  22.2  &  KOI-203 \\     
55701.5588$^\dagger$ &  -25.200  &  0.026  &  -0.019  &  3600  &  16.9  &  KOI-203 \\    
55702.4793$^\dagger$ &  -24.534  &  0.022  &  -0.078  &  3600  &  18.2  &  KOI-203 \\       
55703.4786$^\dagger$ &  -24.732  &  0.035  &  -0.086  &  3600  &  13.2  &  KOI-203 \\       
55704.5938$^\dagger$ &  -25.188  &  0.025  &   0.027  &  2703  &  17.8  &  KOI-203 \\       
55705.5453$^\dagger$ &  -24.590  &  0.013  &  -0.015  &  3600  &  24.5  &  KOI-203 \\    
\hline
\multicolumn{7}{l}{$\dagger$: measurements corrected from Moonlight pollution.} \\
\label{table_rv}
\end{tabular}
\end{table*}

%Thus, the radial velocity variations agree with Doppler shifts caused by a planetary 
%companion, and the transit-signal detected from the Kepler light curve
%could be interpreted as coming from a massive hot jupiter, which we designate as 
%KOI-135b hereafter.
%- Description of the observations carried out with SOPHIE; \\
%- Results from a keplerian fit of the RV measurements; \\
 %- Blend analysis (bisector, mask effects, etc.); \\
%- Anything else?? \\
%\begin{figure}[h]
%\centering
%\includegraphics[width=6.5cm, angle=90]{plot_rv_KO203.ps}
%\caption{To Be Written}
%\label{rv_plot}
%\end{figure}

\subsubsection{Stellar characterisation}
\label{spectral_classification_KOI135}

%\textbf{TBD by Nuno \& Magali} \\
\noindent
Stellar parameters and iron abundances were derived in
LTE, using a grid of plane-parallel, ATLAS9
model atmospheres \citep{Kurucz93} and the 2002
version of the radiative transfer code MOOG
\citep{Sneden73}. The methodology used is
described in detail in \citet{Santosetal04}, \citet{Sousaetal08} and references
therein. The full spectroscopic
analysis is based on the EWs of a set of 49 Fe~{\sc i}
and 13 Fe~{\sc ii} weak lines, by imposing ionization
and excitation equilibrium, as well as a zero
slope between the abundances given by individual
lines and their equivalent width. The errors
in the stellar parameters were derived using
the same methodology as described in \citet{Gonzalez98}.

%The spectra used for this analysis (SOPHIE) were obtained
%in the high-resolution mode (R$\sim$40000). 
The total S/N of
the SOPHIE spectra used for this analysis (co-added from several individual spectra
used to derive radial-velocities) is 36 per pixel (0.02 \AA) at 5500 \AA.
We subtracted the background sky light on the science spectra 
using the second SOPHIE aperture positioned on the sky.
The EWs were carefully measured one by one using the IRAF splot routine.

For FGK dwarfs, the stellar parameters obtained
using this methodology were shown to be compatible
with other estimates in the literature. In particular,
the derived effective temperatures are very close to
the ones obtained using recent applications of the
infra-red flux method \citep{casagrandeetal06}.
For a recent comparison we refer to \citet{Sousaetal11}.

The atmospheric parameters for the host star KOI-135, determined with
the described procedure, are $T_{\rm eff}=6041 \pm 143$~K, log\,$g=4.64 \pm 0.13$, and 
metallicity $[\rm{Fe/H}]=0.33 \pm 0.11$.  No detectable emission was found in the 
cores of Ca ~{\sc ii} H \& K lines. 
The $V \sin{i_{*}}$ of $5.5 \pm 1.5$~\kms ~is in agreement with the rotation period determined
from the light curve, assuming a stellar inclination of $i_{*} \sim 90$~degrees.

\subsection{System parameters}
\label{system_par_KOI135}
%\subsection{Transit Fitting}
To derive the orbital and planetary parameters, 
a simultaneous fit of Kepler photometry and SOPHIE radial
velocity measurements was performed. All the transits were normalised 
by fitting a parabola to the 11~hours intervals of the light curve before 
the ingress and after the egress of each transit.
We discarded two of the thirty-nine available transits, specifically those occurring at 
2455056.139, and 2455089.404~BJD. Indeed, the out-of-transit
intervals around these epochs were not sufficiently wide to correctly normalise 
the transits. 

\begin{figure}[h]
\centering
%\vspace{-2.5cm}
\includegraphics[width=5.5cm, angle=90]{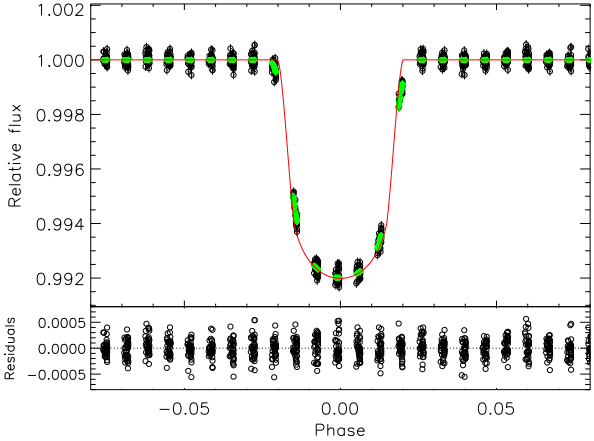}
%\hspace{+1.0cm}
\vspace{+0.5cm}
\caption{
\emph{Top panel}: Unbinned phase-folded transit light curve of KOI-135 (black circles). The red solid line
shows the transit model oversampled five times denser than the Kepler sampling rate. 
The green diamonds indicate the model samples binned to match the Kepler sampling.
\emph{Bottom panel}: the residuals from the best-fit model.}
\label{tr_bestfit_fig_KOI135}
\end{figure}

The nine free parameters of our global best fit
are the orbital period $P$, the transit epoch $T_{\rm tr}$; 
the transit duration $T_{\rm 14}$;
the ratio of the planet to stellar radii $R_{\rm p}/R_{*}$; 
the inclination $i$ between the orbital plane and the plane of the sky;
the Lagrangian orbital elements $h=e~\sin{\omega}$ and  
$k=e~\cos{\omega}$, where $e$ is the eccentricity and
$\omega$ the argument of the periastron; the radial-velocity 
semi-amplitude $K$; and the systemic radial velocity $\gamma_{rel}$.
The two non-linear limb-darkening coefficients $u_{+}=u_{a}+u_{b}$
and $u_{-}=u_{\rm a}-u_{\rm b}$~\footnote{$u_{\rm a}$ and $u_{\rm b}$ are the coefficients
of the limb-darkening quadratic law:
$I(\mu)/I(1)=1-u_{\rm a}(1-\mu)-u_{\rm b}(1-\mu)^2$, where $I(1)$ is the 
specific intensity at the centre of the disc and $\mu=\cos{\gamma}$, 
$\gamma$ being the angle between the surface normal and the line of sight} 
were fixed in our analysis. They can not be let as free parameters
as the orbital period is almost an integer multiple of the long-cadence 
sampling $\delta t =29.42$~min, 
$P=147.986 \cdot \delta t$, which implies that the transit ingress and egress 
are not well sampled (see Fig.~\ref{tr_bestfit_fig_KOI135}). 
The adopted limb-darkening coefficients $u_{\rm a}$ and $u_{\rm b}$ for 
the Kepler bandpass were taken from \citet{Sing10} 
Tables\footnote{ \scriptsize $\rm  http://vega.lpl.arizona.edu/singd/David\_Sing/Limb\_Darkening.html$}, after 
linearly interpolating at the $T_{\rm eff}$, log\,$g$ and metallicity of the star:
$u_{\rm a}=0.375 \pm 0.026$ and $u_{\rm b}=0.277 \pm 0.015$, which 
give $u_{+}=0.652 \pm 0.030$ and $u_{-}=0.098 \pm 0.030$. 

The transit fitting was carried out using the model 
of \citet{Gimenez06, Gimenez09} and a denser temporal 
sampling $\delta t_{\rm model}=\delta t/5$, as suggested by \citet{KippingBakos11}
to overcome the problem of the coarse Kepler sampling (see also \citealt{Kipping10}). 
The $\chi^2$ of each trial model is then computed by binning the model 
samples at the Kepler sampling rate $\delta t$. The search for the best 
solution of our combined fit was performed by using the 
algorithm AMOEBA \citep{Pressetal92} and changing the initial
values of the parameters with a Monte-Carlo method to 
avoid the solution to get stuck into a local minimum.
The 1-$\sigma$ errors of the system parameters were 
estimated through a bootstrap procedure consisting in 
shifting the photometric residuals and, at the same time, 
shuffling the radial velocity ones. During the bootstrap procedure,
the limb-darkening coefficients were let vary within their 
error bars related to the uncertainties of the atmospheric parameters
(see Sect.~6.3 in \citealt{Bouchyetal11} for more details).  

The system parameters and their 1-$\sigma$ errors are listed in 
Table~\ref{starplanet_param_table_135_204}. The radial-velocity 
measurements taken with the SOPHIE spectrograph and 
the solution of the Keplerian fit derived from our combined fit are 
shown in Fig.~\ref{fig_orb_rv}. Figure~\ref{tr_bestfit_fig_KOI135} 
displays the phase-folded transit light curve of KOI-135b and, 
superposed, the transit model.  

As stated before, the transits of KOI-135b are not well sampled because the orbital 
period is almost an integer multiple of the Kepler long-cadence rate. 
For this reason we performed some additional best fits by considering
also transit models with a sampling denser than $\delta t/5$, i.e. 
$\delta t_{\rm model}=\delta t/9$ and $\delta t/15$. We always found
a solution compatible with that reported in 
Table~\ref{starplanet_param_table_135_204} within 1-$\sigma$ and, thus, conclude
that $\delta t_{\rm model}=\delta t/5$ gives an accurate enough solution
for the transit parameters.

%The best-fit parameters were found
%by using the algorithm AMOEBA \citep{Pressetal92} and changing the initial
%values of the parameters with a Monte-Carlo method 
%to find the global minimum of the $\chi^{2}$.
%Following Kipping \& Bakos 2010 (ADD ref.) and Santerne et al. (ADD ref.),
%the transit modelling was performed with a temporal sampling 
%five times denser than that of Kepler, i.e. one point every 5.88~min, 
%and then binning the model samples to 
%match the Kepler sampling rate.
%In such a way, the solution of the transit parameters is more accurate 
%than a simple chi-square minimisation between the Kepler 
%measurements and the transit model (Kipping 2010, ADD ref.). 
%Indeed, the Kepler sampling of 29.4~min 
%inevitably smears all sharp changes occurring in the transit
%ingress and egress (Kipping \& Bakos 2010, ADD ref.).

%%%%%%%%%%%%%%%%%%%%%%%%%%
%%%%%%%%%%%%%%%%%%%%%%%%%%
%%%%%%%%%%%%%%%%%%%%%%%%%%
\begin{table*}
%\vspace{-0.4cm}
\centering
\caption{KOI-135 and 204: planet and star parameters.}            
%\vspace{1cm}
%\begin{minipage}[t]{15.0cm} 
%\setlength{\tabcolsep}{10.0mm}
\setlength{\tabcolsep}{1.0mm}
\renewcommand{\footnoterule}{}                          
\begin{tabular}{l c c}        
\hline\hline                 
%\\
\emph{Fitted system parameters} & KOI-135 & KOI-204 \\
%\multicolumn{3}{l}{\emph{Fitted system parameters}} \\
\hline
Planet orbital period $P$ [days] & 3.024095 $\pm$ 0.000021 & 3.246740 $\pm$ 0.000018 \\ % 0.000007 \\ %_{-0.0008}^{+0.0009}$ \\ %$\pm$ 0.0002 \\
Planetary transit epoch $T_{ \rm tr}$ [BJD-2400000] & 54965.4159 $\pm$ 0.0006 & $54966.3781 \pm 0.0004$ \\ %0.0002$ \\% $\pm$ 0.0012 \\
Planetary transit duration $T_{\rm 14}$ [h] & 2.926 $\pm$ 0.019 & 3.218 $\pm$ 0.043 \\ % _{-0.11}^{+0.10}$ \\ % $\pm$ 0.06 \\
Radius ratio $R_{\rm p}/R_{*}$ & $0.0868_{-0.0007}^{+0.0006}$ & 0.0844 $\pm$ 0.0011 \\ % $\pm$ 0.0014 \\
Inclination $i$ [deg] & $84.35_{-0.40}^{+0.47}$ & $83.78_{-0.55}^{+0.65}$ \\
$u_{+}$~$^a$ & 0.652 $\pm$ 0.030 & 0.668 $\pm$ 0.031 \\
$u_{-}$~$^a$ &  0.098 $\pm$ 0.030 &  0.172 $\pm$ 0.031 \\
$h=e~\sin{\omega}$ &  0.019 $\pm$ 0.021 &  0.006 $\pm$ 0.025 \\
$k=e~\cos{\omega}$ &  -0.016  $\pm$ 0.022 &  0.010 $\pm$ 0.031\\
Orbital eccentricity $e$  & $<0.025$  &  $<0.021$  \\ 
Radial-velocity semi-amplitude $K$ [\ms] & $375 \pm 13$ & $124 \pm 5$ \\ % $\pm$ 0.025 \\
Systemic velocity  $\gamma_{\rm rel}$ [\kms] & $-37.591 \pm 0.007$ & $-35.892 \pm 0.004$ \\ %$\pm$ 0.007 \\
%O-C residuals [\ms] & ?? \\
%Planetary occultation epoch $T_{\rm occ}$ $^a$ [HJD-2400000] & 54276.49 $\pm$ 0.41 \\
%Planetary occultation duration $d_{\rm occ}$ [h] & 2.08 $\pm$ 0.18 \\
%Epoch of periastron $T_{0}$ [HJD-2400000] & 54990.85 $\pm$ 0.08 \\
%& \\
%\\
%\multicolumn{2}{l}{\emph{Derived parameters from radial velocity observations}} \\
%\hline    

%Radial velocity semi-amplitude $K$ [\ms] & 301 $\pm$ 10 \\
%Systemic velocity  $V_{\rm r}$ [\kms] & 15.330 $\pm$ 0.007 \\
%O-C residuals [\ms] & 29 \\
%& \\
%\\
%\multicolumn{2}{l}{\emph{Fitted and fixed transit parameters}} \\
%\hline
%$\theta_{2}$~$^b$ & 0.00483 $\pm$ 0.00009 \\
 %\(\mbox{ \boldmath $\theta_{2}$} \) ~$^b$ & 0.00483 $\pm$ 0.00009 \\
& \\

\multicolumn{3}{l}{\emph{Derived system parameters}} \\
\hline
% $\pm$ 0.04 \\
%Argument of periastron $\omega$ [deg] & $98.9_{-6.8}^{+5.9}$ \\ % $\pm$ 6.4 \\ 
$a/R_{*}$ & $6.81_{-0.20}^{+0.24}$  & $6.45_{-0.26}^{+0.32}$ \\ % $\pm$ 2.15 \\
$a/R_{\rm p}$ & $78.4_{-2.7}^{+3.3}$ & $76.4_{-3.8}^{+4.8}$ \\ % $\pm$ 21 \\
$(M_\star/\Msun)^{1/3} (R_\star/\Rsun)^{-1}$ & $0.773_{-0.022}^{+0.027}$ & $0.699_{-0.028}^{+0.035}$ \\ % $\pm$ 0.09 \\
Stellar density $\rho_{*}$ [$g\;cm^{-3}$] & $0.65_{-0.05}^{+0.07}$ & $0.48_{-0.06}^{+0.07}$ \\ %$\pm$ 0.70\\
Impact parameter $b$ & $0.67_{-0.03}^{+0.02}$ & $0.70_{-0.04}^{+0.03}$ \\ % $\pm$ 0.03\\
& \\

\multicolumn{3}{l}{\emph{Atmospheric parameters of the star}} \\
\hline
Effective temperature $T_{\rm{eff}}$[K] & 6041 $\pm$ 143 & 5757 $\pm$ 134\\
%Spectroscopic surface gravity log\,$g$ [cgs]&  4.5  $\pm$  0.15 &  4.59  $\pm$  0.14 \\
Surface gravity log\,$g$ [cgs] ~$^b$ &  4.26  $\pm$ 0.05  &  4.15  $\pm$ 0.06 \\
Metallicity $[\rm{Fe/H}]$ [dex]& 0.33  $\pm$ 0.11 & 0.26  $\pm$ 0.10\\
Stellar rotational velocity $V \sin{i_{*}}$ [\kms] & 5.5 $\pm$ 1.5 & 4 $\pm$ 2\\
Spectral type &  G0V/G0IV & G2IV  \\
%Spectral type &  ?? & ??  \textbf{with the SpT for the two stars?}) \\
& \\
%& \\
%\\
%
%\\
\multicolumn{2}{l}{\emph{Stellar and planetary physical parameters}} \\
\hline
Star mass [\Msun] ~$^c$ &  1.32 $\pm$ 0.09 &  1.19 $\pm$ 0.10 \\
Star radius [\Rsun] ~$^c$ &  $1.42 \pm 0.07$ &  $1.52 \pm 0.09$ \\
%Photometric surface gravity log\,$g$ [cgs] &  4.25  $\pm$ 0.05  &  4.15  $\pm$ 0.06 \\
Planet mass $M_{\rm p}$ [\Mjup ]  &  $3.23 \pm 0.19$ &  $1.02 \pm 0.07$ \\ %_{-1.08}^{+1.16}$  \\ %  $\pm$ ?? \\
Planet radius $R_{\rm p}$ [\Rjup]  &   $1.20 \pm 0.06 $ &   $1.24 \pm 0.07 $ \\ %_{-0.11}^{+0.12}$  \\ %$\pm$  ?? \\
Planet density $\rho_{\rm p}$ [$g\;cm^{-3}$] &   $2.33 \pm 0.36$  &  $0.65 \pm 0.12$ \\ %_{-2.58}^{+3.47}$ \\ % $\pm$ ?? \\
Planet surface gravity log\,$g_{\rm p}$ [cgs] &  $3.75 \pm 0.04 $ &  $3.21 \pm 0.05 $ \\ %_{-0.07}^{+0.08}$ \\ % $\pm$  ?? \\
Stellar rotation period $P_{*, rot}$ [days]  &  12.9 $\pm$ 0.7 & \\
Age of the star $t$ [Gyr] & $2.8^{+1.0}_{-0.8}$ & $6.95^{+1.1}_{-1.7}$ \\ %$\pm$ 0.09  \\
Distance of the star $d$ [pc] &1950 $\pm$ 250 & 2250 $\pm$ 300  \\ 
%Distance of the star $d$ [pc] & & \textbf{taking also Av into account?} $\pm$ ?? \\
Orbital semi-major axis $a$ [AU] & 0.0449 $\pm$ 0.0010 & 0.0455 $\pm$ 0.0013 \\
%Orbital distance at periastron $a_{\rm per}$ [AU] & 0.141 $\pm$ 0.005 \\
%Orbital distance at apoastron $a_{\rm apo}$ [AU] &  0.179 $\pm$ 0.006 \\
Equilibrium temperature $T_{\rm eq}$ [K] ~$^d$ & 1637 $\pm$ 47 & 1603 $\pm$ 51\\
%Equilibrium temperature at periastron
%$T^{\rm per}_{\rm eq}$ [K] ~$^e$ & $975_{-44}^{+47}$ \\ % $\pm$ 54 \\
%Equilibrium temperature at apoastron
%$T^{\rm apo}_{\rm eq}$ [K] ~$^e$ & $863_{-33}^{+36}$ \\ % $\pm$ 39 \\
%&\\
%\\
\hline       
\multicolumn{3}{l}{$a$: the non-linear limb-darkening coefficients were fixed for the transit fitting but let vary} \\
\multicolumn{3}{l}{within their 1-$\sigma$ error bars during the bootstrap procedure (see text for explanation).} \\
\multicolumn{3}{l}{$b$: from Kepler photometry and evolutionary tracks;} \\
\multicolumn{3}{l}{$c$: from STAREVOL evolutionary tracks;} \\
\multicolumn{3}{l}{$d$: black body equilibrium temperature assuming a uniform heat redistribution to the night-side.} \\
%$c$: \\
%} 
%\multicolumn{3}{l}{$a$: Kepler magnitude from MAST Archive; $b$: from 2MASS catalogue.} \\
%\multicolumn{3}{l}{$a$: Kepler magnitude from MAST Archive; $b$: from 2MASS catalogue.} \\
\end{tabular}
%\end{minipage}
\label{starplanet_param_table_135_204}  
\end{table*}

\subsection{Stellar and planetary parameters}
\label{stellar_planet_param_KOI135}
%\textbf{TBD by Aldo and Magali} \\

\noindent
The stellar density derived from the transit fitting and the STAREVOL evolutionary
tracks \citep{PalaciosPC, Siess06} for the effective temperature and metallicity of KOI-135  
point to a star with mass $M_\star =1.32\pm0.09~\Msun$ and 
radius $R_\star =1.42 \pm 0.07~\Rsun$. The quoted errors
on stellar parameters take into account also the uncertainties inherent in stellar models, 
which are typically 3\% and 5\% for the stellar radius and mass, respectively (\citealt{Southworth11};
see also Sect.~4 in \citealt{desert11}). The latter were quadratically added to the 
statistical errors on $M_\star$ and $R_\star$ related to the stellar density from
the transit best fit.

We point out that the surface gravity deduced from the aforementioned stellar parameters, 
log\,$g=4.26  \pm 0.05$, is compatible with the spectroscopic log\,$g$ at 
2.7~$\sigma$. However, it is well known that the surface gravity derived from 
a spectral analysis is the most uncertain atmospheric parameter and, thus, such an
apparent discrepancy must not be of great concern. In other words, our 
error bars on the spectroscopic log\,$g$ are likely underestimated.
Fixing the log\,$g$ to the photometric
determination and performing again the spectral analysis letting only the $T_{\rm eff}$ and
metallicity vary, changes only slightly the solution reported in 
Sect.~\ref{spectral_classification_KOI135}, well within the 1-$\sigma$ uncertainties. \\

According to the determined stellar parameters, the mass and radius of the hot Jupiter KOI-135b
are $M_{\rm p}= 3.23 \pm 0.19~\Mjup$ and $R_{\rm p}=1.20 \pm 0.06~\Rjup$. The 
corresponding planet density is $2.33 \pm 0.36\, \rm g\;cm^{-3}$. The 
STAREVOL evolutionary tracks indicate that the age of the star, 
and thus of the planetary system KOI-135, is $2.8^{+1.0}_{-0.8}$~Gyr. This age is 
in good agreement with the estimate given by gyrochronology \citep{Barnes07}, 
i.e. $1.6 \pm 0.7$~Gyr~\footnote{using B-V=$0.58 \pm 0.04$ estimated from
Eq.~3 in \citet{SekiguchiFukugita00}}. We point out that the 
agreement is even better when considering 
the ``modified gyrochronology" by \citet{Lanza10}, which takes 
the influence of the planet on the evolution of stellar angular momentum into account 
and gives $3.0 \pm 1.3$~Gyr.

%%%%%%%%%%%%%%%%%%%%%%%%%
%%%%%%%%%%%%%%%%%%%%%%%%%
%%%%%%%%%%%%%%%%%%%%%%%%%
%%%%%%%%%%%%%%%%%%%%%%%%%
%%%%%%%%%%%%%%%%%%%%%%%%%
%%%%%%%%%%%%%%%%%%%%%%%%%
%%%%%%%%%%%%%%%%%%%%%%%%%
%%%%%%%%%%%%%%%%%%%%%%%%%
%%%%%%%%%%%%%%%%%%%%%%%%%
%%%%%%%%%%%%%%%%%%%%%%%%%
%%%%%%%%%%%%%%%%%%%%%%%%%
%%%%%%%%%%%%%%%%%%%%%%%%%
%%%%%%%%%%%%%%%%%%%%%%%%%
%%%%%%%%%%%%%%%%%%%%%%%%%
%%%%%%%%%%%%%%%%%%%%%%%%%
%%%%%%%%%%%%%%%%%%%%%%%%%
%%%%%%%%%%%%%%%%%%%%%%%%%
%%%%%%%%%%%%%%%%%%%%%%%%%

\section{KOI-204}

\subsection{Kepler observations}
\label{Kepler_obs_KOI204}
\begin{figure}[h]
\centering
\includegraphics[width=6.5cm, angle=90]{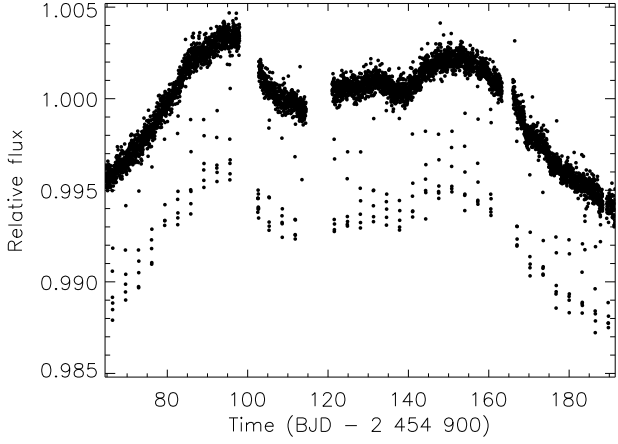}
\caption{The Kepler light curve of KOI-204 with a temporal sampling of 29.4~min
showing 35 transits and long-term variability.}
\label{lcfig_KOI204}
\end{figure}

\noindent
The target KOI-204 is a faint star with Kepler magnitude $\rm K_{p}=14.7$.
Its coordinates, magnitudes and IDs 
are listed in Table~\ref{startable_KOI}.
As for KOI-135, only 
Q1 and Q2 raw light curves were considered.
From these curves the flux excess due to contaminating background stars, 
that is 4.4\% and 6.0\% for Q1 and Q2 data
respectively\footnote{http://archive.stsci.edu/kepler/kepler\_fov/search.php}, 
was subtracted.
%and instrumental trends were appropriately 
%corrected. 
The KOI-204 light curve, displayed in 
Fig.~\ref{lcfig_KOI204}, shows 35 transits with a period of 3.2 days, a
depth of 0.74\%, and a duration of 3~hours. 
Long-term variations with an amplitude of $\sim 1\%$ are also seen in the light curve,
indicating that the star KOI-204 is a slow rotator. 
%although it is not clear to us whether they are of instrumental or astrophysical origin.
After filtering out such variations, the rms of the light curve is $370$~ppm, 
compatible with the median of the errors of the individual 
photometric measurements, i.e. $340$~ppm.

%(\textbf{Rodrigo, Alex, can we conclude on that? How? Any idea on that?}).

%\begin{table}
%\caption{KOI-204 IDs, coordinates, and magnitudes.}            
%\begin{minipage}{3.5 cm} 
%\centering        
%\begin{minipage}[!]{7.0cm}  
%\renewcommand{\footnoterule}{}     
%\begin{tabular}{lcc}       
%\hline\hline                 
%%BJD & RV & $\pm$$1\,\sigma$ & exp. time & S/N p. pix. \\
%%-2\,400\,000 & (km\,s$^{-1}$) & (km\,s$^{-1}$) & (sec) &  (at 550 nm)  \\
%Kepler ID & 9305831 \\
%USNO-A2 ID  & 1350-11449251 \\
%2MASS ID   & 20002456+4545437 \\
%%GSC2.3 ID & NIMR021985 \\
%\\
%\multicolumn{2}{l}{Coordinates} \\
%\hline            
%RA (J2000)  & 20:0:24.55 \\
%Dec (J2000) & 45:45:43.56 \\
%\\
%\multicolumn{3}{l}{Magnitudes} \\
%\hline
%\centering
%Filter & Mag & Error \\
%\hline
%%B$^a$  & 16.68 & 0.14 \\
%%V$^a$  & 15.22 & 0.05 \\
%%B$^a$ & 14.7 & \\ 
%%R$^a$ & 14.1 & \\
%$ \rm K_{p}$$^a$ & 14.68 &  \\ 
%%r$^b$ & 13.88 &  \\
%J$^b$  & 13.34 & 0.03 \\ 
%H$^b$  & 12.97 & 0.02 \\
%K$^b$  & 12.89 & 0.03 \\
%%\\                                    
%%\multicolumn{3}{l}{Proper motion} \\
%%\hline
%%$\mu_{\alpha}$ & 8.0   & ''/yr
%%$\mu_{\alpha}$ & -11.8 & ''/yr
%\hline\hline
%\vspace{-0.5cm}
%%\footnotetext[1]{from USNO-A2.0 catalogue;}
%\footnotetext[1]{Kepler magnitude from MAST Archive;}
%\footnotetext[2]{from 2MASS catalogue.}
%\end{tabular}
%\end{minipage}
%\label{startable_KOI204}      
%\end{table}

%\subsection{Centroid analysis}
%KOI-204 presents a hint of centroid effect with a shift of $0.72 \pm 0.58$~mpx in X 
%and $0.66 \pm 0.54$~mpx in Y (see Fig.~\ref{rain_plot}). 
%In order to confirm that the transits in the Kepler light curve occur on the main target KOI-204 and not on 
%the main contaminant KIC9305838 (Kp=16.6, 8"N from the target) we 
%observed from the ground the transit of KOI-204b during the night of 2011, October the 1st. 
%We used an amateur 14-inch telescope from the Oversky Observatory located at La Palma. 
%Photometric measurements were obtained with a CCD SBIG STL-1001e and using a sloan r' filter. 
%To keep the observed stars in the same pixel during all the night, we used a tip-tilt SBIG AOL-8. 
%We obtained 74 photometric measurement with exposure time of 180s each. 
%Photometric observations were reduced using the free software 
%Munipack\footnote{http://c-munipack.sourceforge.net} and are 
%displayed in Fig.~\ref{KOI204_check}. 
%They clearly show a transit occurring on the star KOI-204 and compatible 
%with those observed by Kepler.

\subsection{Centroid analysis}
KOI-204 presents a hint of centroid effect with a shift of $0.72 \pm 0.58$~mpx in X 
and $0.66 \pm 0.54$~mpx in Y (see Fig.~\ref{rain_plot}). 
According to the MAST database, the main contaminant is the star KIC9305838 with 
Kepler magnitude $K_{\rm p}=16.6$, located
8"~N from KOI-204. In order to confirm that the transits visible in the Kepler light 
curve occur on the main target KOI-204, we performed a photometric follow-up
of this planetary candidate.

\begin{figure}[h]
\begin{center}
\includegraphics[width=\columnwidth]{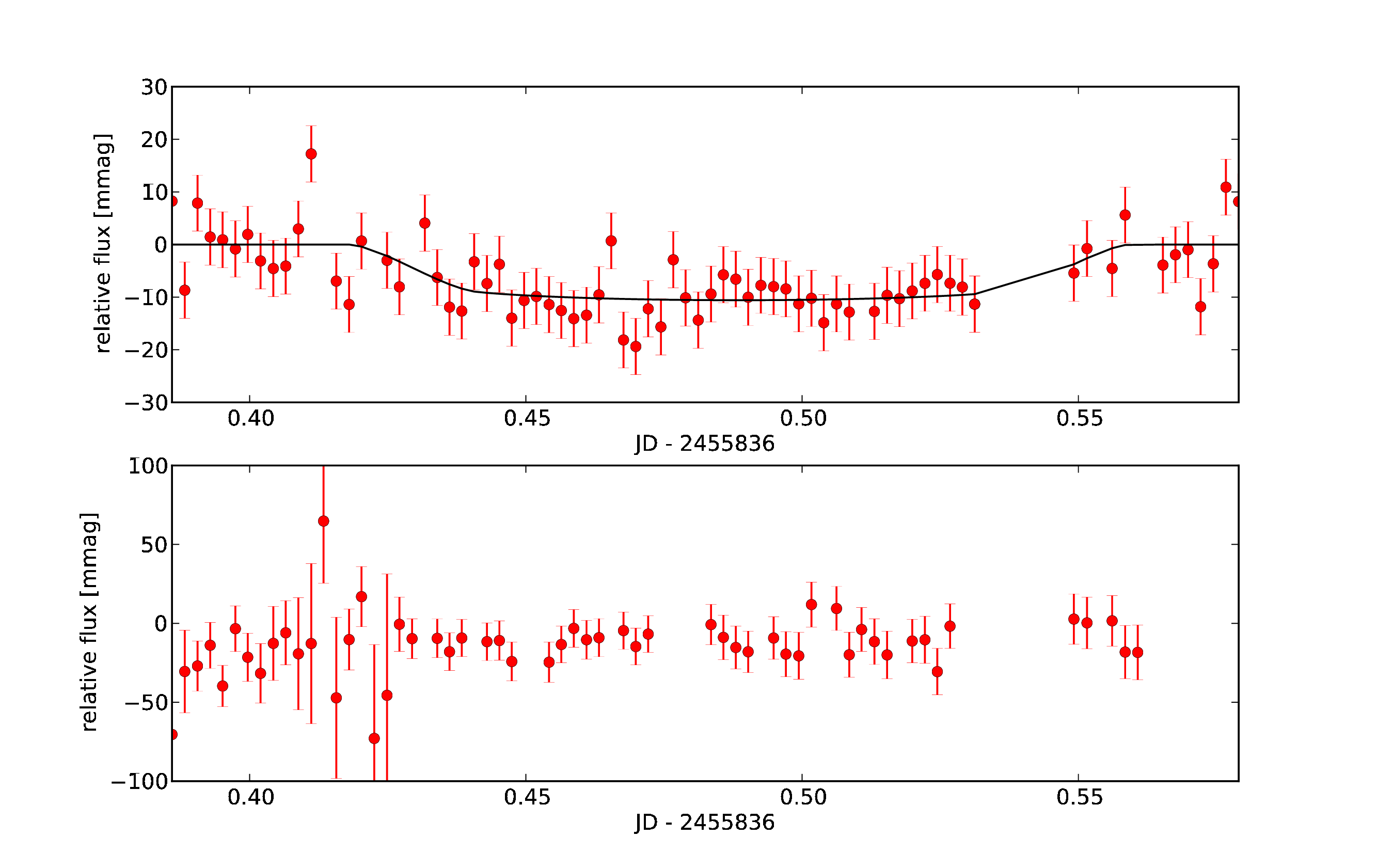}
\caption{Ground-based photometric observation of the transit of KOI-204b. 
\emph{Upper panel}: photometric measurements of the star KOI-204. 
\emph{Lower panel}: photometric measurements of the main contaminant, 
the star KIC9305838. The data clearly show that the transit occurs on the target star KOI-204.}
\label{KOI204_check}
\vspace{-0.5cm}
\end{center}
\end{figure}

%\textbf{TBD by Claire and Alex}

\subsection{Ground-based follow-up}

\subsubsection{Photometric observations}
\label{phot_ground}
The transit of KOI-204b was observed from the ground during the night of 2011, October the 1st, 
with a 14-inch telescope at the Oversky Observatory, La Palma. 
Seventy-four photometric measurements were obtained with a CCD SBIG STL-1001e, 
 using a sloan r' filter and exposure time of 180~s. The angular pixel size is 1.15~arcsec.
To keep the observed stars on the same pixel during all the night, we used a tip-tilt SBIG AOL. 
%We obtained 74 photometric measurements with exposure time of  each. 
Photometric observations were reduced using the free software 
Munipack\footnote{http://c-munipack.sourceforge.net} and are 
displayed in Fig.~\ref{KOI204_check}. 
They clearly show a transit occurring on the star KOI-204
compatible with those observed by Kepler.

\subsubsection{Radial-velocity observations}
Seven radial velocities of KOI-204 were secured with the SOPHIE spectrograph. The 
instrumental mode and measurement extractions were the same as those presented in 
Sect.~\ref{sect_RV_KOI135}.
The full width at half maximum of the fitted CCFs is $11.12 \pm 0.05$~km\,s$^{-1}$, 
and its contrast is $32 \pm 3$~\%\ of the continuum. Moonlight correction was applied 
only to one measurement and is relatively modest (20\,\ms).

The radial-velocity measurements are reported in Table~\ref{table_rv}. The precision
ranges between 16 and 46\,\ms. Figure~\ref{fig_orb_rv} (middle panel) displays them as 
well as the Keplerian fit from the combined photometry and radial-velocity analysis 
(Sect.~\ref{system_par_KOI204}), assuming a circular orbit as the orbital 
eccentricity is $< 0.021$ at 1-$\sigma$.
This fit gives a reduced $\chi^2$ of 0.6 and is thus acceptable, the standard 
deviation of the residuals being $\sigma_{O-C}=9.3$~m\,s$^{-1}$. The 
semi-amplitude of the radial-velocity variation is $K=124\pm5$~m\,s$^{-1}$ 
(see Fig.~\ref{fig_orb_rv}).
%, corresponding 
%to a planet with a mass \mp~$  = 1.02 \pm 0.07$~\Mjup\ 
%(assuming $M_\star = 1.19\pm0.10$\,M$_\odot$).
No radial-velocity drift was detected over the 13-day span of observation.

Radial velocities measured from cross-correlations with F0, K0, or K5 masks give results 
similar to those obtained with the G2 mask, and the CCF bisector spans do not show significant 
variations (Fig.~\ref{fig_bis}, middle panel), so there is no hint for any blend scenario. We can 
thus conclude that the Kepler photometric signal and spectroscopic observations are due to 
a hot Jupiter transiting in front of KOI-204. We designate this new planet as KOI-204b.

%\textbf{TBD by Guillaume.}

%- Description of the observations carried out with SOPHIE; \\

%- Results from a keplerian fit of the RV measurements; \\
 
%- Blend analysis (bisector, mask effects, etc.); \\

%- Anything else?? \\

\subsubsection{Stellar characterisation}
\label{spectral_classification_KOI204}

%\textbf{TBD by Nuno \& Magali} \\

\noindent
To derive the atmospheric parameters, the eight 
SOPHIE spectra, taken to measure the radial-velocity variations, 
were co-added, after subtracting the background sky light from each spectrum.
This allowed us to obtain a single spectrum with a S/N of 25 
per pixel (0.02 \AA) at 5500 \AA. Following the same methodology as described in 
Sect.~\ref{spectral_classification_KOI135}, we determined the 
atmospheric parameters of the host star  
$T_{\rm eff}=5757 \pm 134$~K, log\,$g=4.59 \pm 0.14$, and 
metallicity $[\rm{Fe/H}]=0.26 \pm 0.10$. Moreover,
no emission was observed in the cores of the Ca {\sc ii} H\&K lines,
in agreement with the relatively quiescent light curve of KOI-204.

\subsection{System parameters}
\label{system_par_KOI204}

\begin{figure}[h]
\centering
%\vspace{-2.5cm}
\includegraphics[width=5.5cm, angle=90]{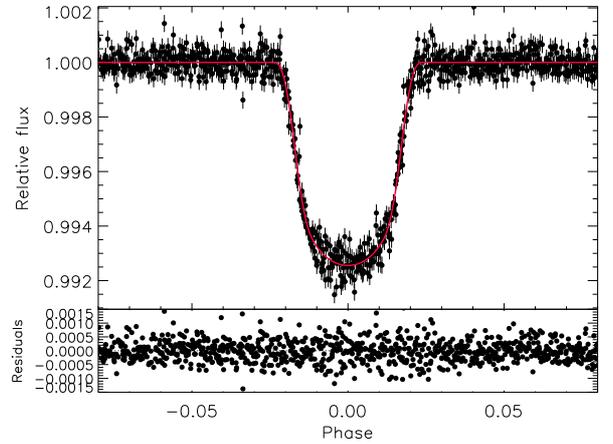}
%\hspace{+1.0cm}
\vspace{+0.5cm}
\caption{
\emph{Top panel}: Unbinned phase-folded transit light curve of KOI-204. The red solid line
shows the transit model rebinned at the Kepler sampling rate (see text for explanation). 
\emph{Bottom panel}: the residuals from the best-fit model.}
\label{tr_bestfit_fig_KOI204}
\end{figure}

\noindent
System parameters and their 1-$\sigma$ errors were derived in the same way
as described in Sect.~\ref{system_par_KOI135}.  They are listed in 
Table~\ref{starplanet_param_table_135_204}. The adopted limb-darkening
coefficients for the Kepler bandpass and the spectral type of KOI-204 are 
$u_{\rm a}=0.448 \pm 0.045$ and $u_{\rm b}=0.227 \pm 0.029$ \citep{Sing10}.
Figure~\ref{tr_bestfit_fig_KOI204} shows the phase-folded transit light curve
and, overplotted, the transit best fit.

\subsection{Stellar and planetary parameters}
\label{stellar_planet_param_KOI204}
%\textbf{TBD by Aldo and Magali} \\

\noindent
The mass and radius of the parent star, derived from the stellar density
given by the transit best fit and STAREVOL evolutionary tracks, are $M_\star =1.19\pm0.10~\Msun$ 
and $R_\star =1.52\pm0.09~\Rsun$. The uncertainties on stellar models
were quadratically added to the statistical errors on the stellar mass and radius.
As for KOI-135, the agreement between the spectroscopic and photometric log\,$g$
is at 3-$\sigma$ level. 

According to the aforementioned stellar parameters, KOI-204b has a radius of $1.24 \pm 0.07~\Rjup$ and a mass of
$1.02 \pm 0.07~\Mjup$, leading to a bulk density of $0.65 \pm 0.12~\rm g\;cm^{-3}$. The age of
the planetary system is estimated to be $6.95^{+1.1}_{-1.7}$~Gyr.

%- Age of the star from isochrones. What about Lithium?

%%%%%%%%%%%%%%%%%%%%%%%%%
%%%%%%%%%%%%%%%%%%%%%%%%%
%%%%%%%%%%%%%%%%%%%%%%%%%
%%%%%%%%%%%%%%%%%%%%%%%%%
%%%%%%%%%%%%%%%%%%%%%%%%%
%%%%%%%%%%%%%%%%%%%%%%%%%
%%%%%%%%%%%%%%%%%%%%%%%%%
%%%%%%%%%%%%%%%%%%%%%%%%%
%%%%%%%%%%%%%%%%%%%%%%%%%
%%%%%%%%%%%%%%%%%%%%%%%%%
%%%%%%%%%%%%%%%%%%%%%%%%%
%%%%%%%%%%%%%%%%%%%%%%%%%
%%%%%%%%%%%%%%%%%%%%%%%%%
%%%%%%%%%%%%%%%%%%%%%%%%%
%%%%%%%%%%%%%%%%%%%%%%%%%
%%%%%%%%%%%%%%%%%%%%%%%%%
%%%%%%%%%%%%%%%%%%%%%%%%%
%%%%%%%%%%%%%%%%%%%%%%%%%

\section{KOI-203 alias Kepler-17}
Although fainter than $\rm K_{p}=14$, the planetary candidate KOI-203 was followed-up
by the Kepler team that, using the High Resolution Spectrometer
(HRS) at the Hobby-Eberly Telescope (HET), detected a radial 
velocity variation of $419.5^{+13.3}_{-15.6} \rm ~m~s^{-1}$, compatible with 
a planetary companion (\citealt{desert11}, hereafter D11). This hot Jupiter, renamed Kepler-17b,
is interesting as it orbits an active solar-like star in 1.486~day
and occults starsposts during transits. Since the stellar rotation period of 11.9~days is
almost an integer multiple of the orbital period, the short-cadence Kepler data (Q4-Q6) 
allowed D11 to see a ``stroboscopic'' effect: the spots are ``mapped'' by the planet
each 45~\ensuremath{^\circ} in longitude. 
Moreover, since the planet occults the same spots for $\sim 100$~days, the true obliquity
of the system, i.e. the angle between the stellar rotation axis and 
the normal to the orbital period, can be estimated close to zero (D11).
Last but not least, the high-quality of Kepler data allowed the Kepler team to detect the 
planetary occultation in the optical with a depth of $58 \pm 10$~ppm and, consequently, 
estimate the geometric albedo of Kepler-17b $A_{\rm g}=0.10 \pm 0.02$. \\

We independently followed up this candidate with the SOPHIE
spectrograph in spring 2011. With eight new SOPHIE spectra, 
we are able to refine the orbital and stellar atmospheric parameters and, thus,
the characterisation of the planetary system.
%Besides that, we confirm the detection of the secondary eclipse in the Kepler 
%light curve and found also the phase variation.
%in brightness as the dayside of the planet rotates
%into and out of view.

%\textbf{TO BE CONTINUED BY ALDO} \\

%From the transit best-fit, using all the 
%available Kepler photometry from Q0 to Q6, and 
%the radial-velocity analysis, Desert et al. 2011 derived the planetary parameters 

\subsection{Kepler observations}
\label{Kepler_obs_KOI203}
\begin{figure}[h]
\centering
\includegraphics[width=6.5cm, angle=90]{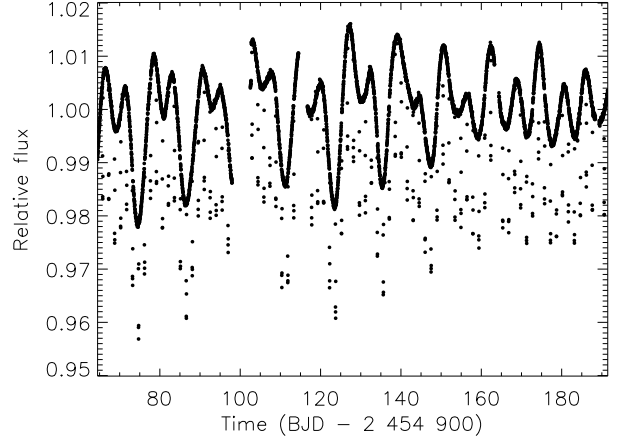}
\caption{The light curve of Kepler-17 with a temporal sampling of 29.4~min
showing 78 transits and variability with a peak-to-peak amplitude of 3\% due to
active regions whose visibility is modulated by the stellar rotation.}
\label{lcfig_KOI203}
\end{figure}

At the time of starting the radial-velocity follow-up, the public Kepler light curve contained  
only Q1 and Q2 data with long-cadence rate. Such light curve, corrected from 
the contamination flux and long-term trends of 
instrumental origin, is shown in Fig.~\ref{lcfig_KOI203}. Stellar variability due to photospheric 
magnetic activity has a peak-to-peak amplitude of $\sim 3\%$.
The long-cadence data, however, do not allow us to see the occultation
of single spots by the planet during the transits, as observed by D11 
with short-cadence data (one point per minute).

%, which prevents from
%observing the above-mentioned ``stroboscopic effect'' and determining 
%the planetary obliquity.

%\begin{table}
%\caption{KOI-203 IDs, coordinates, and magnitudes.}            
%%\begin{minipage}{3.5 cm} 
%\centering        
%\begin{minipage}[!]{7.0cm}  
%\renewcommand{\footnoterule}{}     
%\begin{tabular}{lcc}       
%\hline\hline                 
%%BJD & RV & $\pm$$1\,\sigma$ & exp. time & S/N p. pix. \\
%%-2\,400\,000 & (km\,s$^{-1}$) & (km\,s$^{-1}$) & (sec) &  (at 550 nm)  \\
%Kepler ID & 10619192 \\
%USNO-A2 ID  & 1350-11245067 \\
%2MASS ID   & 19533486+4748540 \\
%GSC2.3 ID & NIMR021985 \\
%\\
%\multicolumn{2}{l}{Coordinates} \\
%\hline            
%RA (J2000)  & 19:53:34.87 \\
%Dec (J2000) & 47:48:54.0 \\
%\\
%\multicolumn{3}{l}{Magnitudes} \\
%\hline
%\centering
%Filter & Mag & Error \\
%\hline
%%B$^a$  & 16.68 & 0.14 \\
%%V$^a$  & 15.22 & 0.05 \\
%%B$^a$ & 14.6 & \\ 
%%R$^a$ & 13.6 & \\
%$ \rm K_{p}$$^a$ & 14.14 &  \\ 
%%r$^b$ & 13.88 &  \\
%J$^b$  & 12.99 & 0.02 \\ 
%H$^b$  & 12.67 & 0.02 \\
%K$^b$  & 12.58 & 0.02 \\
%%\\                                    
%%\multicolumn{3}{l}{Proper motion} \\
%%\hline
%%$\mu_{\alpha}$ & 8.0   & ''/yr
%%$\mu_{\alpha}$ & -11.8 & ''/yr
%\hline\hline
%\vspace{-0.5cm}
%%\footnotetext[1]{from USNO-A2.0 catalogue;}
%\footnotetext[1]{Kepler magnitude from MAST Archive;}
%\footnotetext[2]{from 2MASS catalogue.}
%\end{tabular}
%\end{minipage}
%\label{startable_KOI203}      
%\end{table}

\subsection{Ground-based follow-up}
\subsubsection{Radial-velocity observations}
\label{sect_RV_KOI203}
Eight radial velocities of KOI-203 were secured with the SOPHIE spectrograph in April and May 
2011, with the same instrumental mode and measurement extractions as described in 
Sect.~\ref{sect_RV_KOI135}.
The full width at half maximum of the fitted CCFs is $10.81 \pm 0.12$~km\,s$^{-1}$, 
and its contrast is $34 \pm 4$~\%\ of the continuum. Moonlight correction was applied to all but 
two measurements, but remains always below 25\,\ms.
The radial-velocity measurements are listed in Table~\ref{table_rv}. Their precision
is between 13 and 35\,\ms. Fig.~\ref{fig_orb_rv} (right panel) displays them as 
well as the circular Keplerian fit taking the period and transit epoch given by D11. 
This fit shows an acceptable reduced $\chi^2$ of 1.1 and a standard 
deviation of its residuals $\sigma_{O-C}=17.8$~m\,s$^{-1}$, which indicates
that there is no signature of stellar activity in the radial-velocity residuals.
The derived semi-amplitude of the radial-velocity variation is $K=391\pm11$~m\,s$^{-1}$.
%corresponding to a planet with a mass \mp~$  = 2.28 \pm 0.12$~\Mjup\ 
%(assuming $M_\star = 1.06\pm0.07$\,M$_\odot$).
Here again, blends scenarios can be excluded as cross-correlations performed with 
F0, K0, and K5 masks give similar results to those obtained with the G2 mask, and 
CCF bisector spans are constant  (Fig.~\ref{fig_bis}, right panel). 
We thus concluded in spring 2011 that 
KOI-203b is actually a transiting hot Jupiter. 

Thanks to the radial velocities measured with the HRS spectrograph,
D11 derived a slightly less accurate and larger semi-amplitude 
$K=419.5^{+13.3}_{-15.6}$~m\,s$^{-1}$. Such radial velocities have uncertainties ranging 
between 26 and 199\,\ms, significantly less precise than those obtained with SOPHIE.
The residuals of the HRS Keplerian fit show a 52-\ms\ dispersion 
(D11), to be compared with the $17.8$-m\,s$^{-1}$ dispersion 
obtained after the best fit of the SOPHIE data. 

Our final result is obtained from the combined analysis of the SOPHIE and HRS data (excluding 
the least accurate HRS measurement) and is, however, mainly driven by the SOPHIE data. We  
derived $K=399\pm9$~m\,s$^{-1}$.
%, corresponding to a planet with a mass 
%\mp~$  = 2.33 \pm 0.11$~\Mjup. 
%The uncertainty in the stellar mass is the main source of uncertainty 
%in the planetary mass.
No significant radial-velocity drift was detected in the 275-day span of HET and 
SOPHIE data. 
%nor in the 92-day span of HET data.

\subsubsection{Stellar characterisation}
\label{spectral_classification_KOI203}
%\textbf{TBD by Nuno \& Magali} \\
The SOPHIE spectrum, obtained by co-adding the eight individual spectra 
used for the analysis of radial-velocity variations, was used to derive the 
atmospheric parameters of the host star Kepler-17. Such a spectrum has
S/N$= 31$ per pixel (0.02 \AA) at 5500 \AA, higher than the spectrum
used by D11, i.e. S/N$=17.5$. This allowed us to accurately determine the atmospheric
parameters $T_{\rm eff}$, log\,$g$, and $[\rm{Fe/H}]$. On the contrary,
D11 had to fix the log\,$g$ to the photometric value 
due to their low-quality spectrum (see D11, 
Sect.~3.1, for more details). We found an effective temperature  
slightly hotter than D11, $T_{\rm eff}=5781 \pm 85$~K, although compatible at 1.2~$\sigma$. 
The metallicity $[\rm{Fe/H}]$ and log\,$g$ are in good agreement with the
values reported in D11: $[\rm{Fe/H}]=0.26 \pm 0.10$ and 
log\,$g=4.53 \pm 0.12$. According to our atmospheric parameters,
the B-V of Kepler-17 is $\rm 0.66 \pm 0.03$ \citep{SekiguchiFukugita00},
significantly lower than 0.82 as estimated by D11. As a consequence, the activity estimator
$\log{R^{'}_{HK}}=-4.47$ is higher than the value reported by D11, i.e. -4.61.

\subsection{Stellar and planetary parameters}
\label{stellar_planet_param_KOI203}
The transit fitting was performed in the same manner as for KOI-135b and
KOI-204b (see Sect.~\ref{system_par_KOI135}) and our transit parameters
were in very good agreement with those determined by D11. However,
with much more Kepler data and especially short-cadence data, the error bars of
D11 are significantly lower than ours, as expected. For that reason, 
in Table~\ref{starplanet_param_table_KOI203} we keep only the transit parameters and 
related 1-$\sigma$ errors derived by D11.

From the transit density reported in D11 and STAREVOL evolutionary
tracks for the atmospheric parameters of Kepler-17 
(Sect.~\ref{spectral_classification_KOI203}),
the stellar mass and radius are estimated to be 
$M_\star = 1.16 \pm 0.06~\Msun$ and
$R_\star = 1.05 \pm 0.03~\Rsun$, respectively. 
From our value of $M_\star$ and the radial-velocity 
semi-amplitude recomputed with the new SOPHIE 
measurements (Sect.~\ref{sect_RV_KOI203}), by considering
a circular orbit and the orbital inclination 
given by D11,  the planet mass is
\mp~$  = 2.47 \pm 0.10~\Mjup$. The planet radius derived 
from the stellar radius and the value $R_{p}/R_{*}$ reported in 
D11 is \rp~$=1.33 \pm 0.04~\Rjup$.
All the stellar and planetary parameters are in good agreement with 
those determined by D11 (always within 1~$\sigma$). 
Our slightly larger stellar mass compensates with the 
lower radial-velocity semi-amplitude giving almost the same  
planetary mass as D11. 

The most striking difference concerns the age of the planetary system Kepler-17, 
which, according to our analysis, seems to be younger than 1.8~Gyr 
(at 1~$\sigma$) while D11 found an age of $3.0 \pm 1.6$~Gyr.
The age given by the evolutionary tracks can be compared to that 
from the gyrochronology since the rotation period has been inferred 
from the light curve of Kepler-17: $0.9 \pm 0.2$~Gyr. 
The ``modified gyrochronology'' from
\citet{Lanza10} points to an age of $1.7 \pm 0.3$~Gyr and thus is
more consistent with the extreme limit of the age estimated from evolutionary tracks.

%AGE OF THE SYSTEM FROM ISOCHRONES \
%- Comparison of the age given by the evolutionary tracks with the  
%gyrochronology. The latter gives $t=1050 \pm 420$~Myr or, according to 
%the modified gyrochronology (Lanza 2010), $t=1960 \pm 790$~Myr. \\
%- Can we get an independent estimate of the stellar age from 
%the $R^{'}_{HK}$? This was not done by Desert et al. \\
%- Something else about this planet, internal structure, inflated? \\

\begin{table*}
%\vspace{-0.4cm}
\centering
\caption{KOI-203: planet and star parameters.}            
%\vspace{1cm}
%\begin{minipage}[t]{13.0cm} 
%\setlength{\tabcolsep}{10.0mm}
%\setlength{\tabcolsep}{-4.5mm}
\renewcommand{\footnoterule}{}                          
\begin{tabular}{l c c}        
\hline\hline                 
%\\
\multicolumn{2}{l}{\emph{Fitted system parameters}} \\
\hline
Planet orbital period $P$ [days] & 1.4857108 $\pm$ 0.0000002 & \citet{desert11} \\ %_{-0.0008}^{+0.0009}$ \\ %$\pm$ 0.0002 \\
Planetary transit epoch $T_{ \rm tr}$ [BJD-2400000] & $55185.678035_{-0.000026}^{+0.000023}$ & \citet{desert11} \\ 
Planetary transit duration $T_{\rm 14}$ [h] & 2.276 $\pm$ 0.017 & \citet{desert11} \\ % _{-0.11}^{+0.10}$ \\ % $\pm$ 0.06 \\
Radius ratio $R_{\rm p}/R_{*}$ & $0.130307_{-0.00018}^{+0.00022}$ & \citet{desert11} \\ % $\pm$ 0.0014 \\
Inclination $i$ [deg] & 87.22 $\pm$ 0.15 & \citet{desert11} \\
Quadratic limb darkening coefficient $u_{\rm a}$  & 0.405 $\pm$ 0.007 & \citet{desert11} \\
Quadratic limb darkening coefficient $u_{\rm b}$  &  $0.262^{+0.013}_{-0.015} $& \citet{desert11}\\
%$h=e~\sin{\omega}$ &  $0.120_{-0.024}^{+0.022}$ \\ % $\pm$ 0.01 \\
%$k=e~\cos{\omega}$ &  $-0.019_{-0.012}^{+0.015}$ \\ % $\pm$ 0.01 \\
Orbital eccentricity $e$   &  $<0.001$ & \citet{desert11} \\ 
Radial-velocity semi-amplitude $K$ [\ms] & 399$\pm$ 9  & This work \\ % $\pm$ 0.025 \\
Systemic velocity  $\gamma_{\rm rel}$ [\kms] & $ -24.814  \pm 0.007 $ & This work \\ 
HET-SOPHIE offset   $\gamma_{\rm rel}$ [\kms] & $ -0.019  \pm 0.014 $ & This work \\ 
%O-C residuals [\ms] & ?? \\
%Planetary occultation epoch $T_{\rm occ}$ $^a$ [HJD-2400000] & 54276.49 $\pm$ 0.41 \\
%Planetary occultation duration $d_{\rm occ}$ [h] & 2.08 $\pm$ 0.18 \\
%Epoch of periastron $T_{0}$ [HJD-2400000] & 54990.85 $\pm$ 0.08 \\
%& \\
%\\
%\multicolumn{2}{l}{\emph{Derived parameters from radial velocity observations}} \\
%\hline    

%Radial velocity semi-amplitude $K$ [\ms] & 301 $\pm$ 10 \\
%Systemic velocity  $V_{\rm r}$ [\kms] & 15.330 $\pm$ 0.007 \\
%O-C residuals [\ms] & 29 \\
%& \\
%\\
%\multicolumn{2}{l}{\emph{Fitted and fixed transit parameters}} \\
%\hline
%$\theta_{2}$~$^b$ & 0.00483 $\pm$ 0.00009 \\
 %\(\mbox{ \boldmath $\theta_{2}$} \) ~$^b$ & 0.00483 $\pm$ 0.00009 \\
& \\

\multicolumn{2}{l}{\emph{Derived system parameters}} \\
\hline
% $\pm$ 0.04 \\
%Argument of periastron $\omega$ [deg] & $98.9_{-6.8}^{+5.9}$ \\ % $\pm$ 6.4 \\ 
$a/R_{*}$ & $5.48 \pm$ 0.02 & \citet{desert11} \\
%$a/R_{\rm p}$ & $42.0_{-0.8}^{+0.5}$ \\ % $\pm$ 21 \\
$(M_\star/\Msun)^{1/3} (R_\star/\Rsun)^{-1}$ & 1.001 $\pm$ 0.020 & \citet{desert11} \\
Stellar density $\rho_{*}$ [$g\;cm^{-3}$] & 1.415 $\pm$ 0.084 & \citet{desert11} \\
Impact parameter $b$ & $0.268_{-0.012}^{+0.014}$ & \citet{desert11} \\ % $\pm$ 0.03\\
& \\

\multicolumn{2}{l}{\emph{Atmospheric parameters of the star}} \\
\hline
Effective temperature $T_{\rm{eff}}$[K] & 5781 $\pm$ 85 & This work \\
Surface gravity log\,$g$ [cgs]&  4.53  $\pm$  0.12 & This work \\
Metallicity $[\rm{Fe/H}]$ [dex]& 0.26  $\pm$ 0.10 & This work \\
Stellar rotational velocity $V \sin{i_{*}}$ [\kms] &  6 $\pm$ 2 & This work \\
Spectral type & G2V  \\
& \\

%& \\
%\\
%
%\\
\multicolumn{2}{l}{\emph{Stellar and planetary physical parameters}} \\
\hline
Star mass [\Msun] ~$^a$ &  1.16 $\pm$ 0.06 & This work \\
Star radius [\Rsun] ~$^a$ &  $ 1.05 \pm 0.03$  & This work \\
Photometric surface gravity log\,$g$ [cgs]&  4.46  $\pm$  0.04 & This work \\
Planet mass $M_{\rm p}$ [\Mjup ]  &  $ 2.47 \pm 0.10$ & This work \\ %_{-1.08}^{+1.16}$  \\ %  $\pm$ ?? \\
Planet radius $R_{\rm p}$ [\Rjup]  &   $1.33 \pm 0.04 $  & This work \\ %_{-0.11}^{+0.12}$  \\ %$\pm$  ?? \\
Planet density $\rho_{\rm p}$ [$g\;cm^{-3}$] &   $1.30 \pm 0.14$ & This work \\ %_{-2.58}^{+3.47}$ \\ % $\pm$ ?? \\
Planet surface gravity log\,$g_{\rm p}$ [cgs] &  $3.54 \pm 0.03 $ & This work \\ %_{-0.07}^{+0.08}$ \\ % $\pm$  ?? \\
Stellar rotation period $P_{*, rot}$ [days]  &  11.89 $\pm$ 0.15 & \citet{desert11}\\
Age of the star $t$ [Gyr] & $<1.78$  & This work \\
Distance of the star $d$ [pc] & 800  $\pm$ 100 & This work \\ 
%Distance of the star $d$ [pc] & \textbf{taking also Av into account?} $\pm$ ?? & This work \\
Orbital semi-major axis $a$ [AU] & 0.0268 $\pm$ 0.0005 & This work \\
%Orbital distance at periastron $a_{\rm per}$ [AU] & 0.141 $\pm$ 0.005 \\
%Orbital distance at apoastron $a_{\rm apo}$ [AU] &  0.179 $\pm$ 0.006 \\
Equilibrium temperature $T_{\rm eq}$ [K] ~$^b$ & 1746 $\pm$ 26 & This work \\
Geometric albedo $A_{\rm g}$ & $ < 0.12$ & This work \\
%Equilibrium temperature at periastron
%$T^{\rm per}_{\rm eq}$ [K] ~$^e$ & $975_{-44}^{+47}$ \\ % $\pm$ 54 \\
%Equilibrium temperature at apoastron
%$T^{\rm apo}_{\rm eq}$ [K] ~$^e$ & $863_{-33}^{+36}$ \\ % $\pm$ 39 \\
%&\\
%\\
\hline       
\multicolumn{3}{l}{$a$: from STAREVOL evolutionary tracks;} \\
\multicolumn{3}{l}{$b$: black body equilibrium temperature assuming a uniform heat redistribution to the night-side.} \\
\end{tabular}
%\end{minipage}
\label{starplanet_param_table_KOI203}  
\end{table*}

\subsection{Secondary eclipse and phase variations of KOI-203b}
%\textbf{TBD by Aldo and Alex}
Before the announcement of Kepler-17b by D11, we had independently
detected its secondary eclipse and phase variations, although only with
Q1 and Q2 Kepler data. This was the main reason why, as for KOI-196b 
\citep{Santerneetal11}, we decided to perform a spectroscopic follow-up 
of the Kepler planetary candidate KOI-203b.

\begin{figure}[t]
\vspace{0.1 cm}
\centering
\includegraphics[width=5.5cm, angle=90]{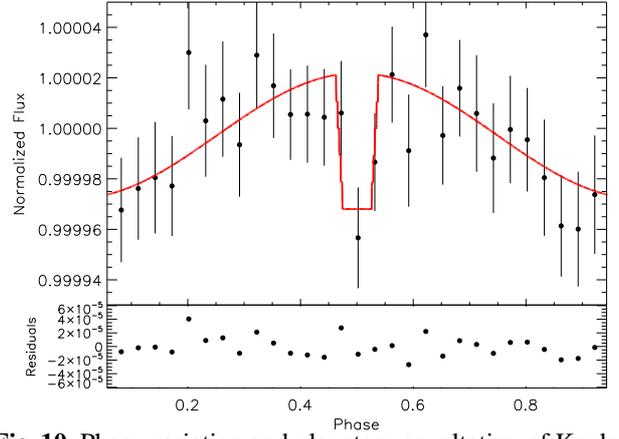}
\caption{Phase variation and planetary occultation of Kepler-17b from
Q1 and Q2 Kepler photometric measurements. 
Data were binned in bins of 0.03 in phase.}
\label{secondary_KOI203}
\end{figure}

To detect both the secondary eclipse and phase variations of 
Kepler-17b, transits were removed from the Q1 and Q2 light curves and the latter 
were divided in subsets of 4 days, approximately one third of the 
stellar rotation period. Each subset was then 
fitted separately with a 4th degree polynomial to remove stellar variability.
A 3-$\sigma$ clipping was applied to this filtered light curve in order to 
get rid of possible outliers. After performing such a de-trending, 
the phase variation and secondary 
eclipse become clearly visible (see Fig.~\ref{secondary_KOI203}).
The rms of the light curve filtered by means of a 4th 
degree polynomial is $260$~ppm in relative flux.
Using a 3rd degree polynomial gives a worse fit of the stellar 
variability, giving a larger rms of $350$~ppm, while, with a 5th
degree polynomial, the rms is equal to $250$~ppm and, thus, comparable
to that obtained with 4th degree polynomials. This motivates our choice
for the use of 4th degree polynomials. In any case, we point out that both
the secondary eclipse and phase variations are visible also using 
3rd and 5th degree polynomials, although quite barely in the first case 
because of the significantly increased rms.

The phase variations and secondary eclipse were modelled 
using Eq.~1 in \citet{Snellenetal09}:

\begin{equation}
F(\phi)=z_{\rm lev} \cdot [1+\sin({\pi \phi})^2 \cdot (1-F_{\rm N/D}) \cdot R_{\rm day}+\psi(\phi)]
\label{equation_Snellen}
\end{equation}
where $\phi$ is the orbital phase, $R_{\rm day}$ is the contrast between the planet day-side flux 
and the stellar flux, i.e. the depth of the secondary eclipse, 
$F_{\rm N/D}$ is the ratio of night-side to day-side flux, and $z_{\rm lev}$ the stellar brightness.
The model of the secondary eclipse $\psi(\phi)$ was computed using the transit model of
\citet{Gimenez06} with no limb-darkening and two free parameters: $R_{\rm day}$ and
the time of the secondary eclipse $T_{\rm sec}$. The duration of the planetary occultation was fixed to that of
the primary transit. 

The simultaneous best fit of the phase variations and secondary transit was carried out 
on the unbinned filtered light curve in the same way as described in 
Sect.~\ref{system_par_KOI135}, i.e. using a Monte Carlo 
method coupled with the downhill simplex algorithm to search for the global minimum of the 
$\chi^2$. As for the transit fitting, we oversampled our model (Eq.~1) with a temporal cadence
five times denser than the Kepler sampling rate.
The error bars of the fitted parameters were estimated with a bootstrap procedure 
that subtracts the best solution to the data, shifts the residuals in time, adds the subtracted 
solution and performs again the best fit. Our $\chi^2$ analysis gives $R_{\rm day}=52 \pm 21$~ppm, 
$F_{\rm N/D} =0$ with an upper limit $F_{\rm N/D} < 0.16$ at $1\sigma$, 
$z_{\rm lev}=0.999973 \pm 0.000011$, and $T_{\rm sec}=2454966.5361 \pm 0.0009$~BJD, 
corresponding to an occultation phase of $0.5003 \pm 0.0006$. 
The best fit is shown in Fig.~\ref{secondary_KOI203} where 
we binned the data in bins of 0.03 with the aim of displaying more distinctly 
the phase variation and the secondary transit.
Our value of $R_{\rm day}$ is perfectly consistent with the occultation depth
found by D11, i.e. $58 \pm 10$~ppm, although our errors are obviously larger
as we have only Q1 and Q2 data. The time of the secondary eclipse allows us to constrain
$-0.0005 < e\cos{\omega} < 0.0014$ at 1-$\sigma$, where $e$ is the eccentricity and $\omega$ the argument of 
periastron. This is compatible with a circular orbit, as also pointed out by D11. For this reason 
we had fixed the eccentricity to zero in the analysis of the radial-velocity observations 
(Sect.~\ref{sect_RV_KOI203}).

By using the more precise occultation depth found by D11,
and a \citet{Kurucz93} model spectrum for the star according
to the atmospheric parameters given in Sect.~\ref{spectral_classification_KOI203},
we estimated a brightness temperature $T_{\rm d, Kepler}=2325^{-50}_{+43}$~K in the 
Kepler bandpass. The planet day-side equilibrium temperature 
ranges from $T_{\rm d}=1746 \pm 26$~K assuming a uniform heat redistribution 
to the night side (or, equivalently, a redistribution factor 
$f=1/4$ in Eq.~1 of \citealt{LopezmoralesSeager07}), 
to $2077 \pm 31$~K for no redistribution ($f=1/2$). For the  
extreme case of a dynamics free atmosphere, where the 
radiative timescale is shorter than the advective timescale, 
$T_{\rm d, max}=2231 \pm 33$~K ($f=2/3$).

\begin{figure}[t]
\vspace{-0.42 cm}
\centering
\includegraphics[width=9.cm, angle=0]{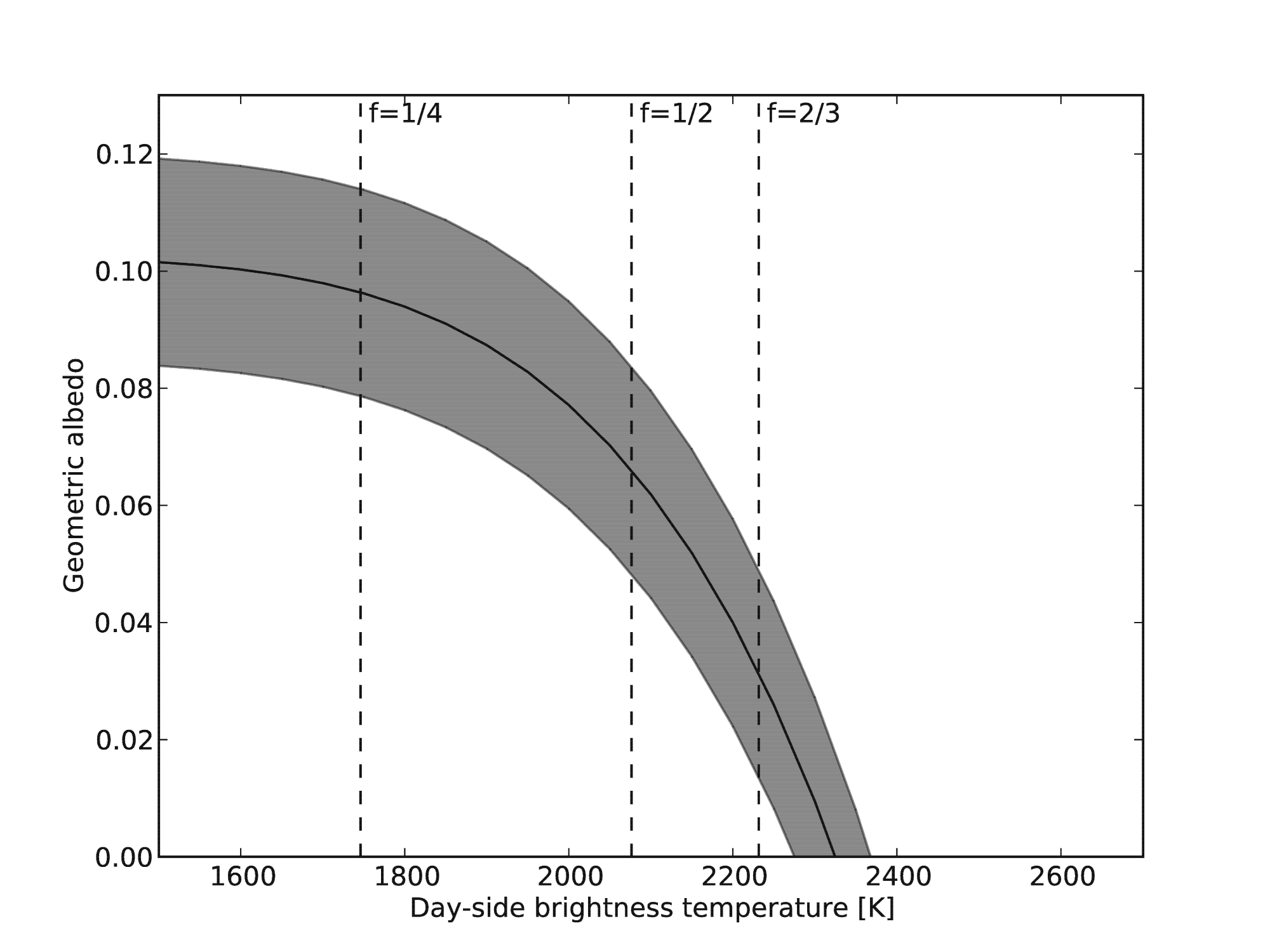}
\caption{Geometric albedo as a function of the planet day-side equilibrium temperature. 
Dashed vertical lines represent the temperature of the planet for i) a perfect heat 
redistribution to the nightside, $T_{\rm d}=1746 \pm 26$~K (left), ii) no thermal circulation 
in the atmosphere, $T_{\rm d}=2077 \pm 31$~K (middle), and iii) a dynamics free
atmosphere where the radiative time scale is shorter than the advective timescale, 
iii) $T_{\rm d, max}=2231 \pm 33$~K (right).
The grey band covers the albedo values allowed by the 1-$\sigma$ 
uncertainty on the occultation depth which is assumed to be 
the dominant uncertainty.}
\label{plot_albedo}
\end{figure}

Fig.~\ref{plot_albedo} shows the geometric albedo $A_{\rm g}$
within the 1-$\sigma$ uncertainty (grey band) as a function of the planet day-side temperature.
The dashed vertical lines indicate the values of $T_{\rm d}$
for the aforementioned redistribution factors $f=1/4, 1/2$ and 2/3. 
As usual, $A_{\rm g}$ is estimated by assuming the planet to radiate as a black body 
and adding a possible reflective component to match the observed occultation 
depth. However, the planet spectrum can significantly differ from a Planck curve 
and have a thermal emission in the optical more important
than expected from a simple black body (e.g., \citealt{Fortneyetal08}; 
\citealt{CowanAgol11} and references therein; see also the discussion by \citealt{Snellenetal10} about the 
optical secondary eclipse of CoRoT-2b). 
Keeping this warning in mind, the geometric albedo, computed as explained above, 
is comprised between 0.01 and 0.12 (see Fig.~\ref{plot_albedo}). 
%The fact that the measured brightness temperature in the Kepler bandpass
%is higher than the maximum allowed temperature $T_{\rm d, max}$ 
%(1.5~$\sigma$) could
%allow for a small fraction of reflected
%Although the measured brightness temperature in the Kepler bandpass is higher than
%the maximum allowed temperature $T_{\rm d, max}$, this is not even
%the proof for the presence of reflected light, for the same reason as above.

Besides detecting the optical occultation in the Kepler light curve,
D11 reported the observations of the Kepler-17b secondary eclipse in the two Spitzer 
band-passes at 3.6 and 4.5~$\mu \rm m$ and derived brightness temperatures of 
$T_{3.6 \mu \rm m}=1880 \pm 110$~K and $T_{4.5 \mu \rm m}=1770 \pm 150$~K (see D11).
If the planet day-side temperature is comprised between the lower and upper limits
of the Spitzer brightness temperatures, 
%$T_{3.6 \mu \rm m}$ and $T_{4.5 \mu \rm m}$,  
the geometric albedo would be $0.06 < A_{\rm g} < 0.12$. However, 
as pointed out by \citet{CowanAgol11}, estimating the planet effective
temperature with only two Spitzer wavebands can lead to errors on the true value 
up to 10\% (see their Fig.~3). Therefore, $T_{\rm d}$ could be higher than the upper limits
given by $T_{3.6 \mu \rm m}$ and $T_{4.5 \mu \rm m}$, which would imply
an even lower albedo, down to $A_{\rm g}=0$ if $T_{\rm d}=T_{\rm d, Kepler}$.

The phase variation we detected in the Kepler light curve including
only Q1 and Q2 data, tells us that, if all the flux from the planet seen in 
the optical is thermal in origin, then only a low fraction of 
stellar heat is transported to the night-side, less than 
$16 \%$ at 1-$\sigma$. In this case, the day-side temperature
would be close to or higher than $\sim 2077$~K. We found no evidence
for a phase shift in the light curve, which is expected in the case of 
no redistribution, under the assumption that the optical occultation comes entirely 
from the thermal flux. 

As suggested by \citet{CowanAgol11}, 
the observation of the thermal phase variation
of Kepler-17b in one of the two IRAC bandpasses, if feasible for such a faint target, would 
in principle permit to estimate the night-side temperature and, thus, 
uniquely determine the circulation efficiency, the planet's albedo and its
day-side temperature.

Finally, we would like to point out that our recomputed activity estimator 
$\log{R^{'}_{HK}}=-4.47$ (Sect.~\ref{spectral_classification_KOI203}), 
higher than the value given by D11, reinforces the propensity for a 
no thermal inversion according to scenario proposed by \citet{Knutsonetal10}
(see also Sect.~7.3 in D11).

\section{Summary and conclusions}

%\textbf{TBD mainly by Aldo and Guillaume but anybody else is welcome} \\
\noindent
We reported the discovery of two new hot Jupiters, namely KOI-135b and KOI-204b, 
and, independently of the Kepler team, confirmed the planetary nature of 
KOI-203b alias Kepler-17b \citep{desert11}, thanks to radial-velocity measurements
performed with the SOPHIE spectrograph at the Observatoire de Haute-Provence.

All the three Jupiter-size planets orbit stars with super-solar metallicity, i.e. $[\rm{Fe/H}] \sim 0.3$~dex.
KOI-135b and KOI-204b have a similar orbital period, $P\sim 3$~days, and 
radius, \rp~$\sim 1.2~\Rjup$, but a different internal structure as their bulk densities 
differ by a factor of 4,  $\rho_{\rm p}=2.33 \pm 0.36$~and $0.65 \pm 0.12\, \rm g\,cm^{-3}$, 
respectively. The density of
Kepler-17b lies in between the two, being equal to the Jupiter's one within the error bars.
The position of KOI-204b and Kepler-17b in the radius vs mass and 
mass vs orbital period diagrams of extrasolar planets
are fairly common, being two slightly inflated planets. 
On the contrary, with a mass of $3.23 \pm 0.19~\Mjup $, KOI-135b reaches the few known transiting planets with short
orbital periods $P < 5$~days and masses between 2.5 and 5 \Mjup. 
Those more similar to KOI-135b, in terms of planetary radius and mass, 
are WASP-32b \citep{Maxtedetal11}
and HAT-P-16b \citep{Buchhaveetal10}.
 
We detected the optical phase variation of Kepler-17b in Q1 and Q2 long-cadence
data and showed that the geometric albedo is not well constrained, being $A_{g} < 0.12$.
If all the optical occultation is thermal, then our modelling of the phase variation 
indicates no redistribution of stellar heat to the night-side with an upper limit of 
16\% at 1-$\sigma$.

Finally, we point out that both Kepler-17 and KOI-135 are interesting targets 
for a detailed study of their photospheric magnetic activity by means
of spot-modelling techniques (e.g., \citealt{Lanzaetal07}). 
%and search for possible  
The Kepler long-term photometry will permit to derive activity cycles
on time-scales of a few years and, more easily, short-term spot cycles 
as found for CoRoT-2b \citep{Lanzaetal09}. In principle, this would allow 
also to search for possible
star-planet magnetic interactions.

%- Say something about a future spot-modelling of the KOI-135 and 203 light curves. \\

\begin{acknowledgements}
We thank the technical team at the Observatoire de Haute-Provence for 
their support with the SOPHIE instrument and the 1.93-m telescope
and in particular the essential work of the night assistants. We are grateful to the
Kepler Team for giving public access to the corrected Kepler light curve and for
publishing a list of good planetary candidates to follow-up. Financial support for
the SOPHIE observations from the Programme National de Plan\'etologie (PNP)
of CNRS/INSU, France is gratefully acknowledged. We also acknowledge 
support from the French National Research Agency (ANR-08-JCJC-0102-01).
A. S. Bonomo acknowledges CNES grant.
NCS acknowledges the support by the European Research Council/European Community 
under the FP7 through Starting Grant agreement number 239953, as well as from 
Funda\c{c}\~ao para a Ci\^encia e a Tecnologia (FCT) through program 
Ci\^encia\,2007 funded by FCT/MCTES (Portugal) and POPH/FSE (EC) and in 
the form of grants reference PTDC/CTE-AST/098528/2008 and PTDC/CTE-AST/098604/2008.
\end{acknowledgements}

~\\
~\\

\end{document}